\pdfoutput=1

\documentclass[11pt,twoside,a4paper,cmspaper,final,collab]{cms-tdr}
\def\svnVersion{297502}\def\cmsCernNo{2013-037}\def\cmsCernDate{\today}\def\cmsMessage{Published in Physics Letters B as \href{http://dx.doi.org/10.1016/j.physletb.2015.07.011}{\doi{10.1016/j.physletb.2015.07.011.}}}

\begin{document}\cmsNoteHeader{EXO-13-007}

\hyphenation{had-ron-i-za-tion}
\hyphenation{cal-or-i-me-ter}
\hyphenation{de-vices}
\RCS$Revision: 297391 $
\RCS$HeadURL: svn+ssh://svn.cern.ch/reps/tdr2/papers/EXO-13-007/trunk/EXO-13-007.tex $
\RCS$Id: EXO-13-007.tex 297391 2015-07-20 17:02:04Z covarell $

\newlength\cmsFigWidth
\ifthenelse{\boolean{cms@external}}{\setlength\cmsFigWidth{0.98\columnwidth}}{\setlength\cmsFigWidth{0.6\textwidth}}
\ifthenelse{\boolean{cms@external}}{\providecommand{\cmsLeft}{top}}{\providecommand{\cmsLeft}{left}}
\ifthenelse{\boolean{cms@external}}{\providecommand{\cmsRight}{bottom}}{\providecommand{\cmsRight}{right}}
\ifthenelse{\boolean{cms@external}}{\providecommand{\cmsTopLeft}{top}}{\providecommand{\cmsTopLeft}{top left}}
\ifthenelse{\boolean{cms@external}}{\providecommand{\cmsTopRight}{middle}}{\providecommand{\cmsTopRight}{top right}}
\newcommand{\thad}{\ensuremath{\tau_\text{had}\xspace}}
\providecommand{\tauh}{\ensuremath{\tau_\mathrm{h}}\xspace}
\newcommand{\te}{\ensuremath{\tau_{\Pe}\xspace}}
\newcommand{\tm}{\ensuremath{\tau_{{\mu}}\xspace}}
\newcommand{\sjrat}{\ensuremath{\tau_{21}}\xspace}
\newcommand{\mprun}{\ensuremath{m_\text{jet}^P\xspace}}
\newcommand{\mzh}{\ensuremath{m_{\Z\PH}}\xspace}
\newcommand{\ciacca}{\ensuremath{c_{\PH}}\xspace}
\newcommand{\givu}{\ensuremath{g_{\mathrm{V}}}\xspace}
\newcommand{\mtt}{\ensuremath{m_{\tau\tau}}\xspace}

\cmsNoteHeader{EXO-13-007}
\title{Search for narrow high-mass resonances in proton-proton collisions at $\sqrt{s} = 8$\TeV decaying to a Z and a Higgs boson}

\date{\today}

\abstract{
A search for a narrow, high-mass resonance decaying into Z and
Higgs (H) bosons is presented.
The final state studied consists of a merged jet pair and a $\tau$ pair resulting
from the decays of Z and H bosons, respectively.
The analysis is based on a data sample of proton-proton collisions at a center-of-mass energy of 8\TeV, collected with the CMS experiment in 2012, and corresponding to an integrated luminosity of 19.7\fbinv.
In the resonance mass range of interest, which extends from 0.8 to 2.5\TeV, the Z and H
bosons are produced with large momenta, which implies that the final products
of the two quarks or the two $\tau$ leptons must be detected within a small
angular interval.
From a combination of all possible decay modes of the $\tau$ leptons,
production cross sections in a range between 0.9 and 27.8\unit{fb}
are excluded at 95\% confidence level, depending on the resonance mass.
}

\hypersetup{%
pdfauthor={CMS Collaboration},%
pdftitle={Search for narrow high-mass resonances in proton-proton collisions at sqrt(s) = 8 TeV decaying to a Z and a Higgs boson},
pdfsubject={CMS},%
pdfkeywords={CMS, physics, resonance production, jets, tau leptons, boosted topology}}

\maketitle

\section{Introduction}

Very recently, the validity of the standard model (SM) of particle physics has been confirmed by the discovery of a Higgs boson with mass near 125\GeV by the ATLAS and CMS experiments~\cite{Aad:2012tfa,HIGGS1}. Though the SM successfully describes a broad range of high energy phenomena, the solution to remaining problems with the structure of the SM, particularly the hierarchy problem, leads naturally to the introduction of physics beyond the standard model (BSM),
possibly at the \TeVns{} scale~\cite{ArkaniHamed:1998rs, PhysRevLett.49.970, Barbieri:1982eh, Agashe:2004rs, PhysRevD.75.055014, PhysRevLett.83.3370}.
Many of
the BSM models predict the existence of heavy resonances with masses of the order of a \TeVns{},
which may have sizable couplings to the gauge and Higgs boson fields of the
SM~\cite{Zprime1,Zprime2,Composite1,Composite2}.
We consider here one important family among these models,
which incorporate composite Higgs bosons~\cite{Composite1,Composite2}. In these models,
the Higgs boson is a pseudo-Nambu-Goldstone
boson of a broken global symmetry.
Other composite bound states beyond the Higgs boson are expected
to exist and could be experimentally observed.

Several searches for massive resonances decaying into pairs of vector bosons or Higgs
bosons have been performed by the ATLAS and CMS experiments~\cite{Aad2015163, Aad:2014pha, Aad:2014xka, Aad:2014yja, CMS:ZZWW, Khachatryan:2014hpa, Khachatryan:2014xja, RECENTAtlas, Khachatryan:2015lba, Khachatryan:2015yea, Aad:2015owa, Aad:2015uka}.
In this analysis, we search for a resonance with a mass in the range 0.8--2.5\TeV
decaying to ZH, where the Z boson decays to $\Pq \Paq$ and the Higgs boson decays to $\tau^+\tau^-$.
It is assumed that the natural width of the resonance is negligible in comparison to the
experimental mass resolution, which is between 6\% and 10\% of the mass of the resonance,
depending on the mass.
There is also a small variation with the type of decay channel because of the dependence of the resolution on the number of neutrinos in the final state.
In the model considered, the spin of the resonance is assumed to be one.
However, it has been verified that the analysis is insensitive to the
angular distributions of the decay products and therefore applies to other
spin hypotheses.

The theoretical model used as benchmark in this work is described
in Ref.~\cite{Pappadopulo:2014qza}.
In this model a heavy $SU(2)_L$ vector triplet (HVT) containing neutral (\cPZpr) and charged ($\PW^{'\pm}$)
spin-1 states is introduced.
This scenario is well-motivated in cases where the
new physics sector is either weakly coupled~\cite{Schmaltz:2010xr}, or
strongly coupled, e.g., in the minimal composite model~\cite{Bellazzini:2012tv}.
The cross sections and branching fractions ($\mathcal{B}$) for the heavy triplet model depend
on the new
physics scenario under study and can be characterized by three parameters in the
phenomenological Lagrangian: the strength of the couplings to fermions
$c_\mathrm{F}$, to the Higgs $\ciacca$, and the self-coupling $\givu$.
In the case of a strongly coupled sector, the new heavy resonance has larger
couplings to the W, Z, and H bosons, resulting in larger branching fractions for
the diboson final states. Our benchmark model characterizes this scenario
by choosing the parameters $\givu=3$ and $c_\mathrm{F}=-\ciacca=1$, which configure a strongly coupled sector.

In the high-mass case under study, the directions of the particles stemming from Z and H boson decays are separated by a small angle. This feature is referred to as the ``boosted'' regime.
For the case of $\Z \to \Pq \Paq$, this results in the presence of one single reconstructed jet after hadronization called a ``Z-jet''.
The novel feature of this
analysis is the reconstruction and selection of a $\tau$ pair in the boosted regime. The presence of
missing energy in $\tau$ decays does not allow a direct determination of the invariant mass.

In the following, we label $\tau$ decays in a simplified way: $\tau^{\pm} \to \Pe^{\pm} \nu \PAGn$ as ``\te'', $\tau^{\pm} \to \mu^{\pm} \nu \PAGn$ as ``$\tm$'',
and $\tau^{\pm} \to (n \pi)(m \PK)\nu$ as ``$\tauh$'', where $n$ and $m$
can be 0, 1, 2, or 3, and the pions and kaons can be either charged or neutral.
Six channels, depending on the combinations of $\tau$ decays, are studied separately
and labeled as all-leptonic (\te\te, \te\tm, \tm\tm), semileptonic ($\te\tauh$, $\tm\tauh$), and all-hadronic ($\tauh\tauh$).

The experimental strategy is to reconstruct and identify the two bosons and to combine their information into a variable that can discriminate between signal and background and on which a statistical study can be performed. This variable is the estimated mass of the $\cPZpr$ after applying dedicated reconstruction techniques to the boosted $ \Pq \Paq$ and $\tau\tau$ pairs (\mzh).
The \mzh distribution would show an excess of events at the assumed $\cPZpr$ mass if a signal were present.

\section{CMS detector}

A detailed description of the CMS detector, together with a definition
of the coordinate system used and the relevant kinematic variables, can be found in
Ref.~\cite{Chatrchyan:2008zzk}.
The central feature of the CMS detector is a 3.8\unit{T} superconducting
solenoid of 6\unit{m} internal diameter.  Within the field volume are the
silicon tracker, the crystal electromagnetic calorimeter (ECAL), and
the brass and scintillator hadron calorimeter (HCAL).
The muon detectors are located outside the solenoid and are installed
between the layers of the steel flux-return yoke of the solenoid.
In addition, CMS has extensive forward calorimetry, in
particular two steel and quartz-fiber hadron forward calorimeters.

\section{Data sample and simulation}

The analysis is based on a data sample collected
by the CMS experiment in proton-proton collisions at a center-of-mass energy of 8\TeV in 2012, corresponding to an integrated luminosity of 19.7\fbinv.
Events are selected online
by a trigger that requires the presence of at least one of the following:
either a hadronic jet reconstructed by the anti-\kt
algorithm~\cite{Cacciari:2008gp} with a distance parameter of 0.5,
transverse momentum \pt larger than 320\GeV, and $\abs{\eta}<5.0$;
or a total hadronic transverse energy, \HT, defined as the scalar sum
of the transverse energy of
all the jets of the event, larger than 650\GeV. The transverse energy
of a jet is defined as the reconstructed energy multiplied by the sine of
the polar angle of the jet axis.
Using events selected by less restrictive, pre-scaled triggers, it has
been verified that the
efficiency of this trigger after applying the offline event selection is above 99\%.  The difference from
100\% is considered as a systematic uncertainty.

The process $\Pq \Paq \to\PZpr\to\Z\PH \to \Pq \Paq \tau^+\tau^-$ is simulated
at parton level using a \MADGRAPH 5 1.5.11~\cite{Alwall:2011uj} implementation of the model described in Ref.~\cite{ref_SignalModel}.
Seven signal samples are generated with masses between 0.8 and 2.5\TeV.
For this mass interval, the $\cPZpr$ production cross section times
branching fraction to ZH ranges from 179.9 fb ($m_{\cPZpr} = 0.8\TeV$) to
0.339 fb ($m_{\cPZpr} = 2.5\TeV$).
Although the main sources of background are estimated using observed events, Monte Carlo (MC)
simulations are used to develop and validate the methods
used in the analysis. Background samples are generated using
\MADGRAPH 5 1.3.30 (Z/$\gamma$+jets and W+jets with leptonic decays), \POWHEG 1.0 r1380 (\ttbar and single
top quark production)~\cite{Nason:2004rx,Frixione:2007vw,Alioli:2010xd,Alioli:2011as},
and \PYTHIA 6.426~\cite{Sjostrand:2006za} (SM diboson production and QCD multijet events with large \HT). Showering and
hadronization are performed with \PYTHIA and $\tau$ decays are simulated using
\textsc{tauola} 1.1.5~\cite{TAUOLA} for all simulated samples.
\GEANTfour~\cite{Agostinelli:2002hh} is used for the simulation of the CMS detector.

\section{Event reconstruction}

A particle-flow (PF) algorithm~\cite{CMS-PAS-PFT-09-001, CMS-PAS-PFT-10-001} is used to identify and to reconstruct
candidate charged hadrons, neutral hadrons, photons, muons, and electrons produced in
proton-proton collisions. Jets and $\tauh$ candidates are then reconstructed
using the PF candidates.
The jet energy scale is calibrated through correction factors that
depend on the \pt and $\eta$ of the jet.
These factors were computed using a data set of proton-proton collisions at $\sqrt{s} = 8$\TeV, corresponding to an integrated luminosity of 19.7\fbinv,
following the method described in~\cite{CMS-JME-10-011}.
All particles reconstructed with the PF algorithm are used to determine the missing
transverse momentum, \ptvecmiss. In first approximation,
\ptvecmiss is defined as the negative vector
sum of transverse momenta of all reconstructed particles~\cite{CMS-JME-12-002}.

Jets are reconstructed using the Cambridge--Aachen (CA)
algorithm~\cite{Wobisch:1998wt}, with a distance parameter of 0.8,
chosen so that it contains the hadronization products of the two quarks from the Z boson.
Jet pruning and subjet-searching algorithms are applied to these jets as
in Ref.~\cite{CMS:ZZWW}.
In these algorithms the original jets are re-clustered by removing
pileup and underlying-event particles at low-\pt~and large angle.
The term pileup refers to additional interactions occurring in the same LHC bunch crossing.
We define \mprun~as the invariant mass of the jet constituents after the
pruning procedure.
This invariant mass provides good discrimination
between Z-jets and quark/gluon-jets since it tends to be shifted towards the energy
scale at which the jet was produced.
We also define a quantity called ``N-subjettiness", $\tau_N$, that is sensitive to the
different jet substructure characteristics of quark/gluon and Z-jets, as~\cite{Thaler:2010tr}:
\begin{equation}\label{eq:subjettiness}
\tau_N = \frac{1}{d_0}\sum_kp_{\mathrm{T},k}\min(\Delta R_{1,k},\Delta
R_{2,k},\ldots,\Delta R_{N,k}),
\end{equation}
where $N$ is the number of subjets in which the original jet can
be reclustered with the \kt
algorithm~\cite{Catani:1993hr,Ellis:1993tq};
the index $k$ runs over the PF constituents of the jet; $p_{\mathrm{T},k}$ is the transverse momentum of the $k$th constituent;
$\Delta R_{n,k}$ is a distance defined as $ \sqrt{\smash[b]{(\Delta\eta_{n,k})^2+(\Delta\phi_{n,k})^2}}$
where $\Delta\eta_{n,k}$ and $\Delta\phi_{n,k}$ are the differences in pseudorapidity and azimuthal angle
between the $k$th constituent and the $n$th subjet axis; and $d_0=\sum_kp_{\mathrm{T},k}R_0$ is
a normalization factor with $R_0$ equal to the original jet distance parameter.
The variable $\tau_N$ quantifies the tendency of a jet to be composed of $N$ subjets, having smaller values for jets with a $N$-subjets-like configuration.
We define \sjrat as the ratio between the 2-subjettiness and the
1-subjettiness, $\sjrat = \tau_2/\tau_1$.
The variables \mprun~and \sjrat~have been shown to have a good discrimination power
between signal and background~\cite{Khachatryan:2014vla}, therefore in the following
they are used to define signal and background enriched regions of the analysis.

In order to match trigger requirements and avoid inefficiencies close to the
threshold, at least one jet in the event is required to have
$\pt > 400$\GeV and $\abs{\eta} < 2.4$.
In addition, this jet is required to pass
minimal consistency requirements on the fraction of charged and neutral particles contributing to it, to avoid fake jets from isolated noise patterns in the calorimeters or the tracker systems.
While the CA jet selection
is common to all the channels considered, the reconstruction of the $\tau\tau$
system is performed differently depending on the $\tau$ decay channel.

The all-leptonic channels are identified by combinations of electrons, muons,
and \ptvecmiss,
which are products of the decay of a pair of $\tau$ leptons from
the Higgs boson.
Electrons are reconstructed by combining the information from an ECAL energy cluster
with that of a matching track in the silicon tracker~\cite{Baffioni:2006cd}. Electrons are selected if they have $\pt > 10$\GeV,
 $\abs{\eta} < 2.5$, and satisfy requirements on the ECAL shower shape, the ratio of energies measured
in HCAL and ECAL around the electron candidate, the compatibility with the
primary vertex of
the event~\cite{vertexes}, and the track-cluster matching parameters.
Muon candidates~\cite{Chatrchyan:2012xi} are reconstructed by performing a global track fit in
which the silicon tracker and the muon system information is combined.
For the \tm\tm~channel, to avoid identification inefficiencies caused by the small angular separation
of the two muon trajectories,
the second muon candidate is
reconstructed with a different algorithm in which tracks in the silicon
tracker are matched in space to signals in the muon detectors~\cite{CMS:ZZWW}.
Muons are required to have $\pt > 10$\GeV, $\abs{\eta} < 2.4$ and to pass
additional requirements on the quality of the track reconstruction, on
the impact parameter of the track, and on
the number of measurements in the tracker and the muon systems.
Electron and muon candidates are required to satisfy particle-flow based isolation
criteria that require low activity in a cone around the lepton, the isolation cone,
after the removal of particles due to additional interactions.
Because the lepton from the other signal $\tau$ decay in the boosted pair can fall
in the isolation cone, other electrons and muons are not considered in the computation
of the isolation criteria.

In the semileptonic channels, a lepton selected with all the criteria above is combined
with a $\tauh$ candidate. The reconstruction of $\tauh$ starts from
the clustering of jets using the anti-\kt algorithm
with a distance parameter of 0.5. Electrons and muons, identified by
looser criteria than the nominal
ones used in the analysis, are removed from the list of particles used
in the clustering if they fall within the jet distance parameter.
The $\tauh$~is reconstructed and identified using the
``hadron-plus-strips'' technique~\cite{CMS-PAS-TAU-11-001},
which searches for the most common decay modes of the $\tauh$
starting from charged hadrons and photons forming $\pi^0$ candidates.
We select $\tauh$ candidates with $\pt > 20$\GeV and $\abs{\eta} < 2.3$.
Electrons and muons misidentified as $\tauh$ are suppressed using dedicated
criteria based on the consistency between the measurements in the tracker,
the calorimeters, and the muon detectors. Finally, loose
PF-based isolation criteria are applied to the $\tauh$ candidates, not
counting electrons and muons in the cone.

In the all-hadronic $\tau\tau$ channel, a subjet-searching technique~\cite{Butterworth:2008iy}
is applied to all CA-jets (distance parameter $R=0.8$) in each
event to identify the $\tauh$ candidates.
At the next-to-last step of the clustering
algorithm, there are two subjets, which are ordered by mass. If both have $\pt > 10$\GeV and the mass of the
leading subjet is smaller than $2/3$ of the mass of the original merged jet, the two objects are used
as seeding jets for $\tau$ lepton reconstruction via the ``hadron-plus-strips'' technique. If any of the criteria above fail,
the procedure for one of the subjets is performed again for a maximum of four iterations.
The efficiency for finding subjets with this method
in signal events is 92\%, independent of \pt, for $\tauh$ with $\pt > 40$\GeV.
In the lowest bin investigated (\pt between 20 and 40\GeV) the efficiency is around 80\%.

The visible mass, $m_\text{vis} $, of the $\tau\tau$ system
is defined as the invariant mass of all detectable products of the two decays.
Because the unobserved neutrinos can carry a significant fraction of the $\tau\tau$
energy/momenta, this variable is not suited for reconstructing resonances that include
the $\tau\tau$ system among its decay products.
Instead, the Secondary Vertex fit (SVFIT) algorithm described
in~\cite{Chatrchyan:2014nva}, which combines
the \ptvecmiss with the
visible momenta to calculate a more precise estimator of the kinematics of the
parent boson, is used to reconstruct the $\tau\tau$ system in all search
channels.

\section{Background composition}
The composition of the background remaining after reconstruction is different for each of the search channels.

In the \te\te, \te\tm, and \tm\tm~channels, the background is almost
entirely composed of Z/$\gamma$+jets events with genuine $\tau$ or other
lepton decays.
In the $\te\tauh$ and $\tm\tauh$ channels, additional significant contributions
to the total background
come from W+jets and \ttbar events with leptonic W-boson decays, and a hadronic
jet misidentified
as $\tauh$. Among \ttbar events, those
with one W boson decaying leptonically and one decaying
to quarks can potentially produce a signal-like structure in \mprun~and \sjrat.
We refer to this as the ``\ttbar peaking contribution'' in the following.

The background in the $\tauh\tauh$~channel is dominated by
QCD multijets production. There is a small but
non-negligible contribution from Z+jets,
W+jets, and \ttbar production. For all these
processes, it is possible that genuine $\tauh$ or at least one
extra jet or lepton
misidentified as $\tauh$ allow the event to pass the selection.

In all channels there is a very small, irreducible component of genuine
SM dibosons, which are not distinguishable from signal, except for the
non-peaking structure in \mzh.

\section{Event selection}

In all channels, the boosted Z boson
decaying to $\Pq \Paq$ is identified by requiring the
selection: $70 < \mprun < 110$\GeV and $\sjrat < 0.75$.
This region is referred to as the ``signal region''.

In the all-leptonic and semileptonic channels, the $\tau\tau$ four-momentum
 estimated from SVFIT is combined
with that of the CA-jet to obtain
the resonance mass \mzh.
Several preselection requirements are
applied to
remove backgrounds from low-mass resonances and from overlaps of lepton
and $\tau$ lepton reconstruction in the detector:
$m_\text{vis} > 10$\GeV, $\Delta R_{\ell\ell}> 0.1$ (where
$\Delta R = \sqrt{\smash[b]{(\Delta\eta)^2 + (\Delta\phi)^2}}$ and $\ell$ denotes electrons, muons, or hadronically decaying taus), $\abs{\ptvecmiss} > 20$\GeV,
and $p_{\mathrm{T}, \tau\tau} > 100$\GeV, as estimated from the SVFIT
procedure.

Since the background in the all-hadronic channel is initially dominated by QCD
multijet events, a different preselection is
applied for the all-hadronic channel.
Only events that have not been included in the
all-leptonic or semileptonic categories are considered in this category.
The event is then separated into two hemispheres containing the decay products of the
two bosons by requiring the following preselection:
$\abs{\ptvecmiss} > 40$\GeV, $\abs{\Delta\phi(\text{CA-jet},\tauh)} >2.0$ and
$\abs{\Delta\phi(\ptvecmiss,\tauh)} <1.5$, for each of the two $\tauh$ candidates.

Further criteria investigated for signal selection in all channels
include tighter requirements on variables like
the \pt of the highest-\pt (leading) lepton
or $\tauh$ and $m_{\tau\tau}$ as estimated from
the SVFIT
procedure. An upper limit is placed on $\Delta R_{\ell\ell}$
in order to reject W+jets events, where a jet misidentified as a $\tau$ lepton is
usually well-separated in space from the isolated lepton. The number of
b jets in the event also provides a useful criterion to reduce the \ttbar
contribution. Jets may be identified as b jets, using the combined secondary vertex
algorithm~\cite{Chatrchyan:2012jua} which exploits observables related to
the long lifetime of b hadrons, and are considered if not overlapping with $\tau$ candidates and CA-jets.
Those b jets are clustered with the anti-\kt jet algorithm,
with a distance parameter $R$ = 0.5.
Optimization of the selection on these variables is based on
the Punzi factor of merit ($\mathcal{P}$)~\cite{punzi}, defined as:
$\mathcal{P} = \varepsilon_{\text{sig}}/(1 + \sqrt{B})$, where $\varepsilon_{\text{sig}}$
is the signal efficiency and $B$ is the background yield after applying the selection.
The results of the
optimization are listed in Table~\ref{tab:optim}. It has been verified that
these results
are not sensitive to the choice of \mzh~window used
to evaluate $\varepsilon_{\text{sig}}$ and $B$.
In Table~\ref{tab:effbkg} we show the efficiency of
the selection in signal events for all search channels.

\begin{table*}[!ht]
\centering
\topcaption{Summary of the optimized event selection for the six $\tau\tau$ channels. The selection variables are explained in the text. The
label $\ell$ refers to electrons, muons, and $\tau$ leptons decaying hadronically.}\label{tab:optim}
\begin{tabular}{l|ccc}
\hline
Selection  & $\te\te$, $\te\tm$, $\tm\tm$ & $\te\tauh$, $\tm\tauh$ & $\tauh\tauh$ \\ \hline
$\abs{\ptvecmiss}$       &   $>$100\GeV           & $>$50\GeV             & $>$80\GeV     \\
$p_{\mathrm{T},\ell}^\text{leading}$& --- & $>$35\GeV             & $>$50\GeV     \\
$N_\text{b-tagged jet}$ & $=0$      & $=0$                & ---          \\
$\Delta R_{\ell\ell}$ & $<$1.0   & $<$1.0            & $<$1.0   \\
$m_{\tau\tau}$        & ---         & ---                  & 105--180\GeV \\ \hline
\end{tabular}
\end{table*}

\section{Background estimation}

Because of the non-uniformity of the background composition, different
estimation techniques are used in each channel.

In the \te\te, \te\tm, and \tm\tm~channels the main background source
lacks events with a genuine massive boson decaying
to quarks, therefore a technique based on sidebands
of the \mprun~and \sjrat~variables
is used for background estimation. In an enlarged search region
defined by $\mprun > 20$\GeV, we define
the ``sideband region'', inverting the selections on \mprun and $\sjrat$, therefore
including both $\mprun$ regions outside the signal range and regions with $\sjrat > 0.75$.

\begin{figure}[hbtp]
\centering
\includegraphics[width=0.48\textwidth]{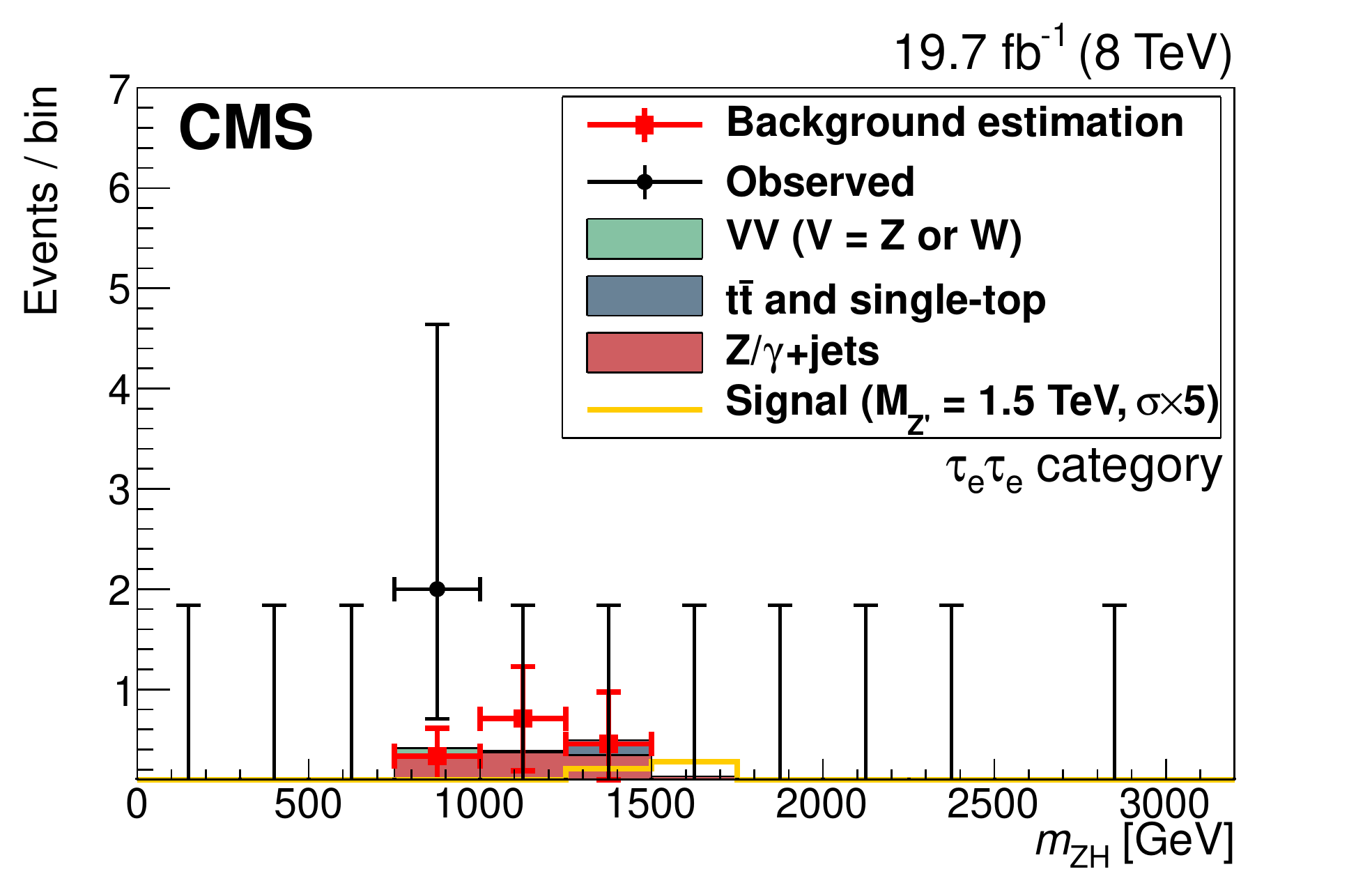}
\includegraphics[width=0.48\textwidth]{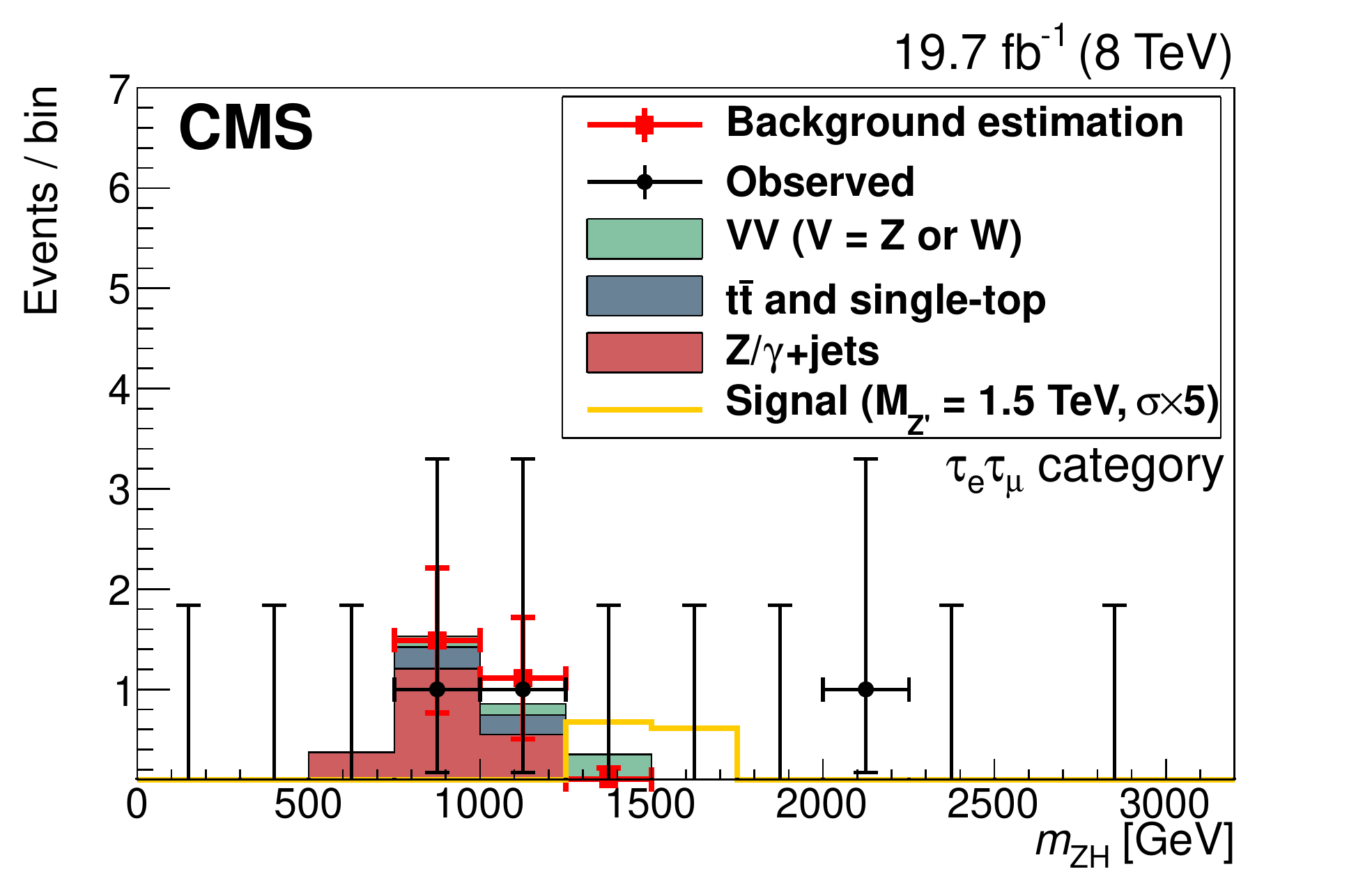}
\includegraphics[width=0.48\textwidth]{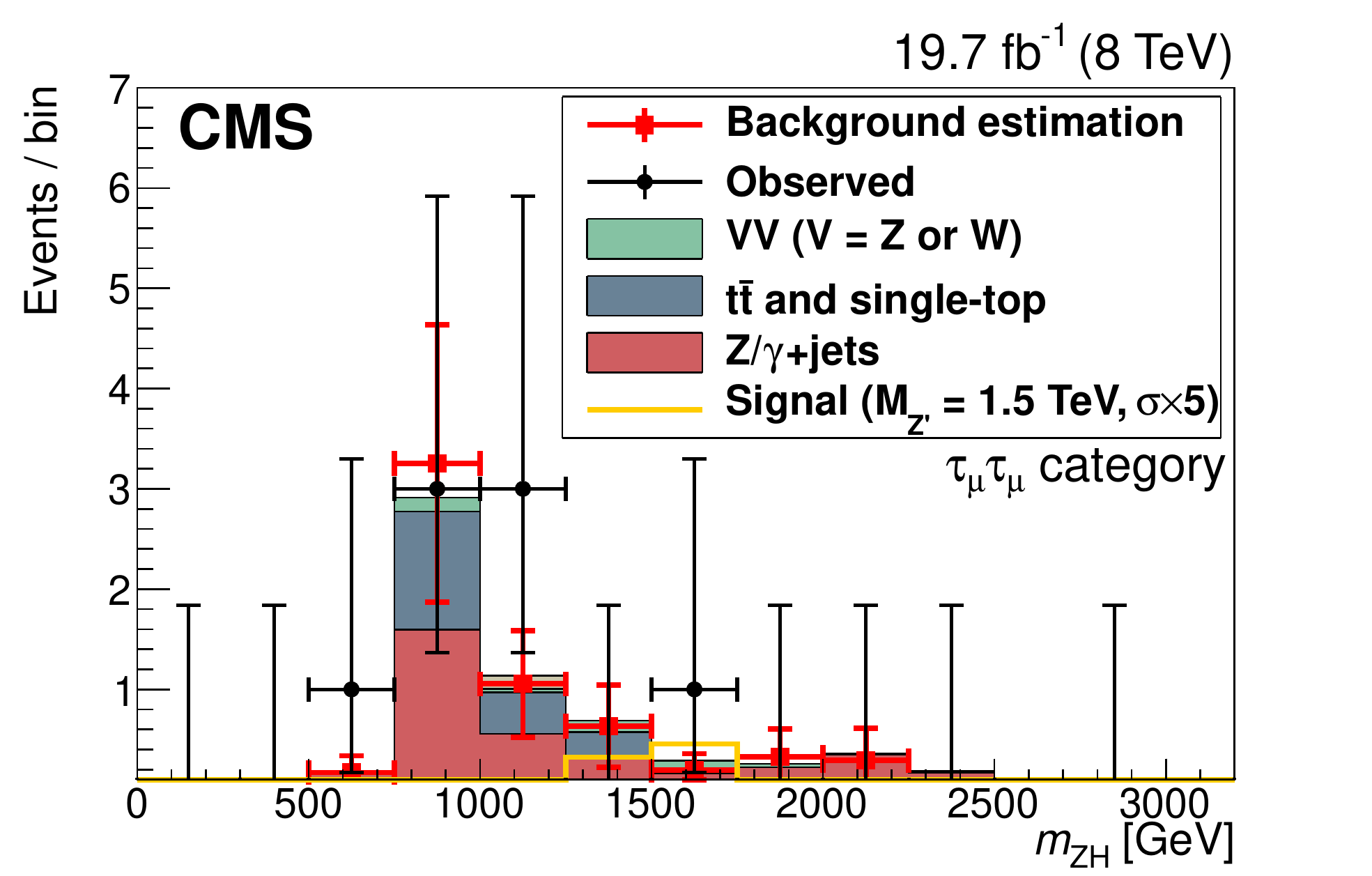}
\caption{Observed
distributions of \mzh for the all-leptonic channels along with the
corresponding MC expectations for signal and background, as well as
background estimation derived from data: (\cmsTopLeft) \te\te~category; (\cmsTopRight) \te\tm~category;
(bottom) \tm\tm~category. Ten equal-size histogram bins cover the region
from 0 to 2.5\TeV, while a single bin is used at higher \mzh~because of the
limited number of MC and data events.
The signal cross section is scaled by a factor of 5.}
\label{fig:yields1}

\end{figure}

The total background is estimated in intervals of \mzh, using the formula:
\begin{equation}
\label{eq:alphaRatioMethod}
N_\text{bkg}(x) = \mathcal{N} \: N_\mathrm{sb}(x) \: \alpha(x),
\end{equation}
where $x = \mzh$, $\mathcal{N}$ is a normalization factor, $N_\mathrm{sb}(x)$ is the number of
events observed in the sideband region, in bins of \mzh,
and $\alpha(x)$ is a binned ratio between the shapes of the \mzh~distributions
in the signal and sideband region, taken from the sum of MC components.
The normalization factor is found through a fit of the observed pruned jet mass
distribution, following the procedure used
in Ref.~\cite{CMS:ZZWW}. The pruned jet mass distribution in the region
$20 < \mprun < 200$\GeV, $\sjrat < 0.75$ is fit in MC samples
with the following function:
\begin{equation}
\label{eq:fitFunc}
F(x) = \mathcal{N} \:\re^{ax} \: \bigl(1 + \erf[(x-b)/c]\bigr),
\end{equation}
where ``$\erf$" is the error function and the parameters $a$, $b$ and $c$
are estimated from the MC simulation.  A fit to the observed distribution,
excluding the signal region, is then used to determine $\mathcal{N}$.
 Figure~\ref{fig:yields1} shows the observed
distributions of \mzh~in all-leptonic channels, along with the
corresponding MC expectations for signal and background, as well as
the background estimation derived with the above procedure.

\begin{figure}[hbtp]
\centering
\includegraphics[width=0.48\textwidth]{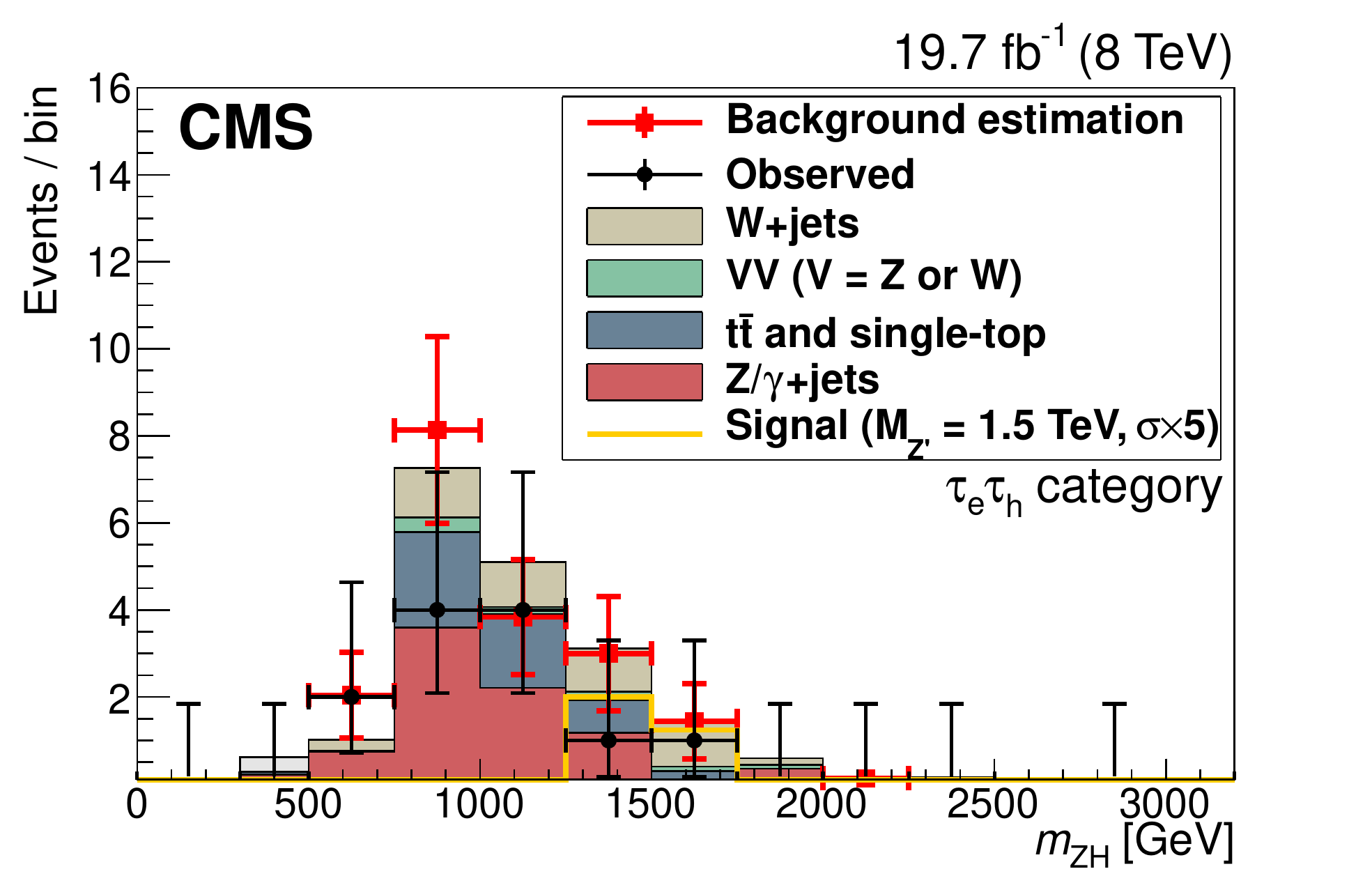}
\includegraphics[width=0.48\textwidth]{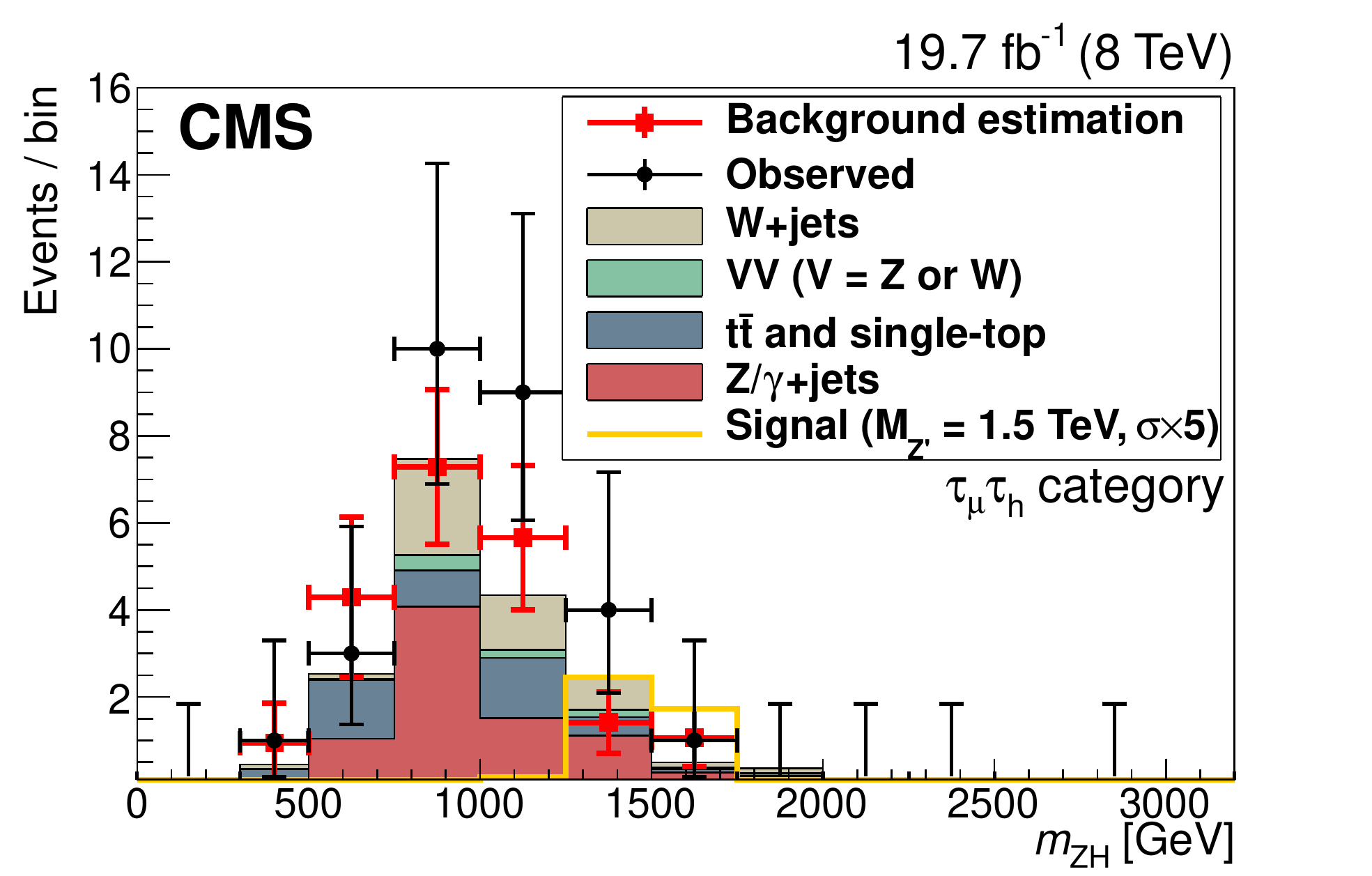}
\caption{Observed
distributions of \mzh for the semileptonic channels along with the
corresponding MC expectations for signal and background, as well as
background estimation derived from data: (\cmsLeft) $\te\tauh$~category; (\cmsRight) $\tm\tauh$~category. Ten equal-size histogram bins cover the region
from 0 to 2.5\TeV, while a single bin is used at higher \mzh~because of the
limited number of MC and data events.
The signal cross section is scaled by a factor of 5.}
\label{fig:yields2}

\end{figure}

In the semileptonic channels, a control sample
defined by the preselection described before, but
requiring at least one b-tagged jet, is selected. It has been
established with simulation that
more than 95\% of this sample is composed of \ttbar events.
Two scale factors (SFs)
relating the ratio of the observed to simulated event rates,
one for the \ttbar peaking
contribution and the other for the \ttbar combinatorial background,
are estimated from this control sample.
The pruned jet
mass distribution is fit with the sum of two functions:
\ifthenelse{\boolean{cms@external}}{
\begin{multline}
F_{\ttbar}(x) = N(\text{non-peaking}) \: \re^{Ax}
\: \bigl(1 + \erf[(x-B)/C]\bigr) +\\ N(\text{peaking}) \: \mathcal{G}(D,E)
\label{eq:fitFunc2}
\end{multline}
}{
\begin{equation}
F_{\ttbar}(x) = N(\text{non-peaking}) \: \re^{Ax}
\: \bigl(1 + \erf[(x-B)/C]\bigr) + N(\text{peaking}) \: \mathcal{G}(D,E)
\label{eq:fitFunc2}
\end{equation}
}
where $A$, $B$, and $C$ define the shape of the non-peaking
component, analogous to Eq.~(\ref{eq:fitFunc}), and $\mathcal{G}(D,E)$ is a
Gaussian function of mean $D$ and standard
deviation $E$. The values of these two parameters are fixed to those found
in the analysis searching for vector boson pair resonances~\cite{CMS:ZZWW}
because we are using the same Z-jet reconstruction.
From this fit, the two scale factors between data and MC are found,
one for each contribution:
$r^{\mathrm{SF}}_1 = N(\text{peaking})_\text{data}/$ $N(\text{peaking})_\mathrm{MC}$
and
$r^{\mathrm{SF}}_2 = N(\text{non-peaking})_\text{data}/N(\text{non-peaking})_\mathrm{MC}$.
The same procedure as for the all-leptonic channels is then applied,
fitting the observed sideband distribution but using a modified function,
given by the sum of the \ttbar contribution and the function of
Eq.~(\ref{eq:fitFunc}),
where the \ttbar normalization is fixed at the MC expectation, scaled by the
two SFs.
Figure~\ref{fig:yields2} shows the
distributions of \mzh~in semileptonic channels, along with the
corresponding MC expectations and
the background estimation derived with the above procedure.

\begin{figure}[hbtp]
\centering
\includegraphics[width=0.48\textwidth]{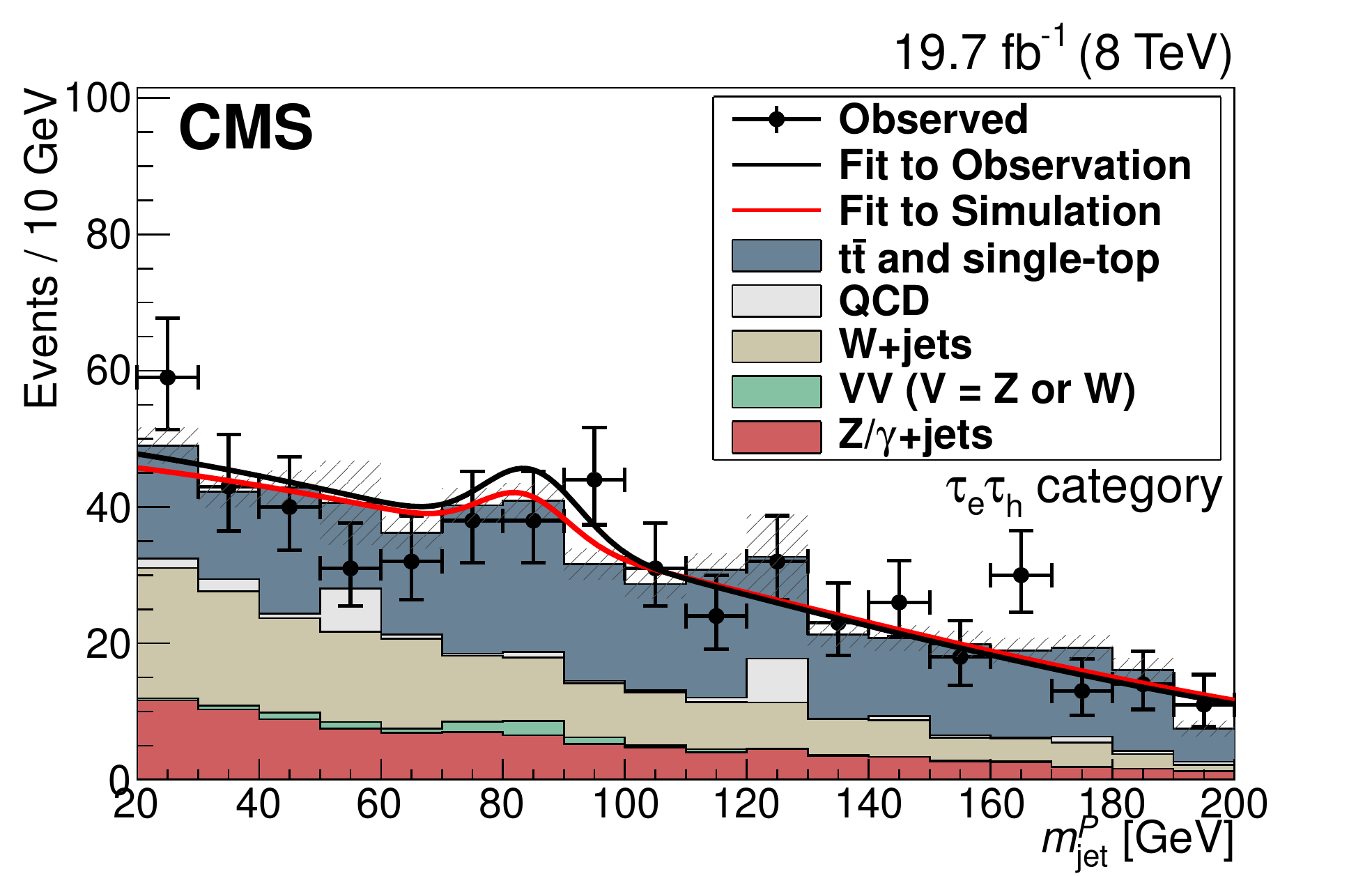}
\includegraphics[width=0.48\textwidth]{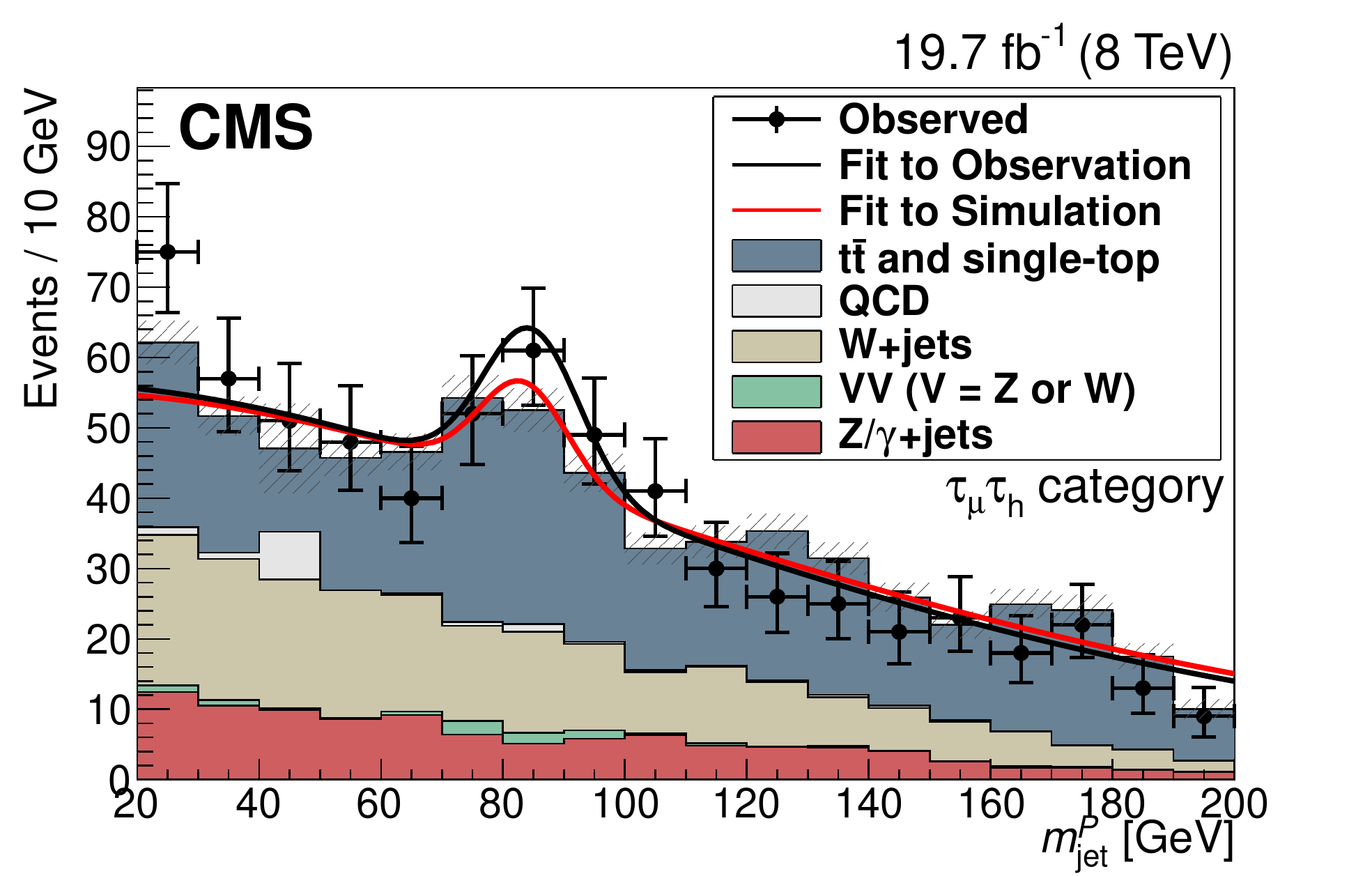}
\caption{Observed
distributions of \mprun~for the semileptonic channels along with the
corresponding MC expectations for signal and background: (\cmsLeft) $\te\tauh$~category; (\cmsRight) $\tm\tauh$~category.
Fits are performed for MC and data (as discussed in the text).}
\label{fig:yields2presel}

\end{figure}

For each of the methods used, consistency checks comparing data and background
predictions are performed using samples of events at the preselection level,
that are expected to have small contributions from potential signal resonances.
In the case of the semileptonic channels, we show in
Fig.~\ref{fig:yields2presel} the distribution of \mprun~for data and MC
at the preselection level. The black line, representing the fit to data,
is obtained by the sum of
Eqs.~(\ref{eq:fitFunc}) and (\ref{eq:fitFunc2}), with the \ttbar shape as
obtained from the control sample, the \ttbar normalization is
fixed to MC scaled by the two SFs, and the other components are free in the
sideband fit. An overall agreement between data and prediction is observed.
The background prediction in the signal region is 156 $\pm$ 26
events, with an observation of 151 events, for the $\te\tauh$ channel and
$204 \pm 31$
events, with an observation of 203 events, for the $\tm\tauh$ channel.

\begin{figure}[hbtp]
\centering
\includegraphics[width=0.5\textwidth]{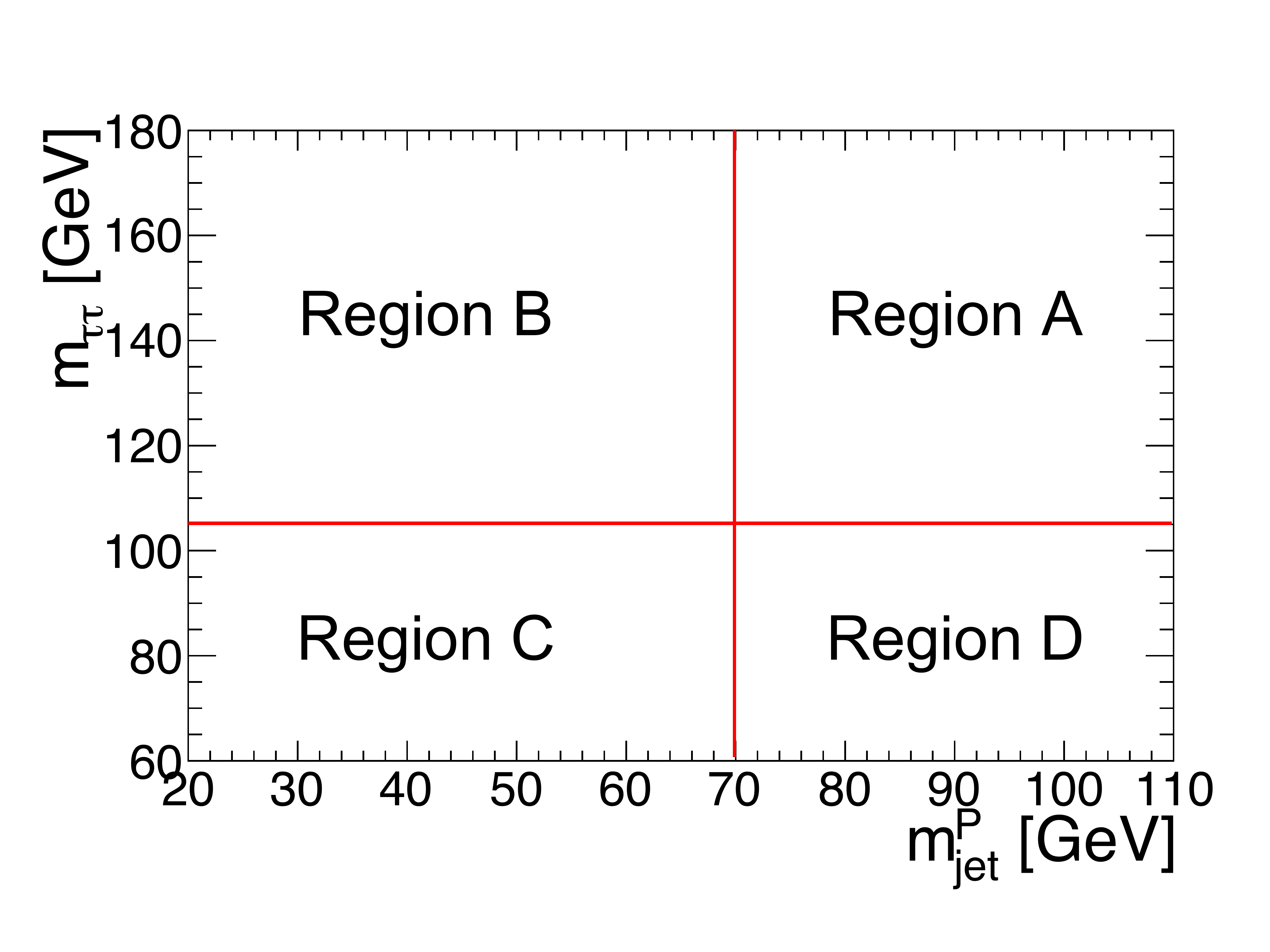}
\caption{Definitions of the A, B, C, and D regions in the \mprun~/ \mtt~plane
used in the background estimation for the all-hadronic channel.}
\label{fig:abcd}

\end{figure}

In the all-hadronic channel, for events where the leading jet
satisfies the requirement $\sjrat < 0.75$, a plane is defined using the
\mprun~and \mtt~variables and four regions are considered, as
shown in Fig.~\ref{fig:abcd}.
Most of the signal events are expected in region A, while regions B, C, D are
dominated by background events.
Studies of the correlation factors for simulated events and
in regions orthogonal to the signal region show that
the variables $\mprun$ and $\mtt$ are essentially uncorrelated.
 In this case,
the total number of background events in the region A can be estimated as:
\begin{equation}
\label{eq:ABCDmethod}
N_\text{bkg} = (N_\mathrm{B} \: N_\mathrm{D}) / N_\mathrm{C}.
\end{equation}
The method described by Eq.~(\ref{eq:ABCDmethod}), called ``ABCD method'',
gives a background prediction
in the signal region that has been checked to be insensitive to possible
signal contamination in the regions B, C, D.

\begin{figure}[hbtp]
\centering
\includegraphics[width=\cmsFigWidth]{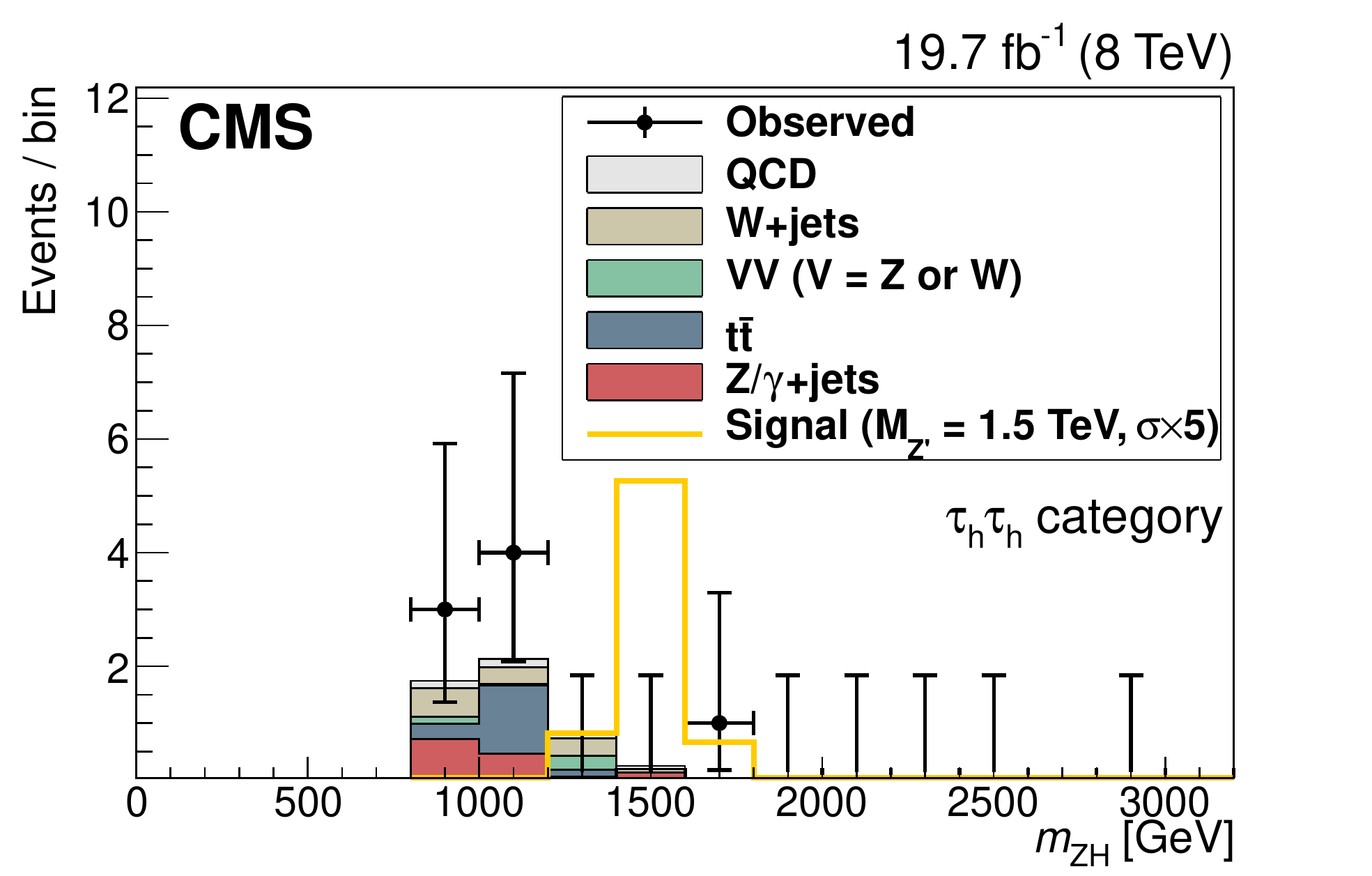}
\caption{Observed
distributions of \mzh for the $\tauh\tauh$ category along with the
corresponding MC expectations for signal and background. Ten equal-size histogram bins cover the region
from 0 to 2.5\TeV, while a single bin is used at higher \mzh~because of the
limited number of MC and data events.
The signal cross section is scaled by a factor of 5.
}
\label{fig:yields3}

\end{figure}

Figure~\ref{fig:yields3} shows the observed
distributions of \mzh~in the $\tauh\tauh$~channel, along with the
corresponding MC expectations for signal and background.
The low number of events in regions B, C, D is not sufficient
to derive the shape of the distribution in the signal region using the ABCD method.
We use the results from this method to compute the cross section upper limits, which
are obtained without assumptions about the shape of the distributions.
The ABCD method is checked using an alternative
background estimation technique, where \ttbar, W+jets and Z+jets background
contributions are given by Eq.~(\ref{eq:alphaRatioMethod}), while the QCD
multijet background is estimated from a control sample of events where
at least one
$\tau$ candidate fails the isolation requirement. The same control sample
is used
to obtain the shape of the QCD distribution in the signal region presented in
Fig.~\ref{fig:yields3}.

\section{Systematic uncertainties}

The sources of systematic uncertainty in this analysis, which affect either
the background estimation or the signal efficiencies, are described below.

For the signal efficiency, the main uncertainties come from
 the limited number of signal MC events
(3--10\%), the integrated luminosity
(2.5\%)~\cite{LumiUncert}, and
the uncertainty on the modeling of pileup
(0.2--2.2\%). Hereafter, the
ranges indicate the different channels and mass regions used in the evaluation
of the upper limits.
The  scale factors for lepton identification are derived from
dedicated analyses of observed and simulated $\cPZ\to \ell^{+}\ell^{-}$
events, using the ``tag-and-probe" method~\cite{Chatrchyan:2012xi,CMS-PAS-EGM-10-004,CMS-PAS-TAU-11-001}.
The uncertainties in these factors are taken as systematic uncertainties and amount to
1--4\% for
electrons, 1--6\% for muons and 9--26\% for $\tau$ leptons decaying hadronically.
The jet and lepton four-momenta are varied over a range given by the
energy scale and resolution uncertainties~\cite{CMS-JME-10-011}.
In this process, variations in the lepton and jet four-momenta
are propagated consistently to \ptvecmiss.
For the all-leptonic and semileptonic channels,
additional uncertainties come from
the procedure of removing nearby tracks
and leptons used in the hadronic $\tau$ reconstruction, and
from the isolation variable computation in the case of boosted topologies.
The inefficiency resulting from these procedures, as measured in signal
simulation, is assigned as a systematic uncertainty, corresponding to
1--16\% for $\tau$ reconstruction and 1--21\% for isolation.
In the all-hadronic analysis, a constant uncertainty of 10\% is assigned
for the application of the $\tau$ reconstruction procedure to collimated
subjets, comparing the performance for isolated and non-isolated $\tau$ leptons in simulation.
The jet trigger efficiency has an uncertainty of $<$1\%, as determined
from a less selective trigger.
Following the method derived for vector boson identification in merged jets~\cite{CMS-JME-13-006},
a scale factor of $0.94 \pm 0.06$ is used
for the efficiency of the pruning and subjet searching techniques applied
on the CA jet,
where the uncertainty is included in the estimation of the overall systematic uncertainty.
For the b tagging, data-to-MC corrections derived from several control
samples are applied and the uncertainties
on these corrections are propagated as systematic uncertainties in the analysis (2--6\%).
The procedure used to derive the b-tagging systematic uncertainties is described in Ref.~\cite{Chatrchyan:2012jua}.

The uncertainties in the background estimate are dominated by the limited numbers of MC events
and sideband data events (4--16 events in all-leptonic channels, 34--37 events in semileptonic
channels and 29 in the all-hadronic channels).
In the analysis of the all-leptonic and
semileptonic channels, additional uncertainties in the background yields
of 10--96\% originate from the limited number of events of the background MC
samples used in the computation of the $\alpha(x)$ quantity, and 18--47\%
from the normalization fit.

\section{Results}

Table~\ref{tab:effbkg} shows the
signal efficiencies (computed using
a sample generated with corresponding $\tau$ decays),
the background expectation and the number of observed events
 for the six analysis channels.

\begin{table*}[!htb]
\centering
\topcaption{Summary of the signal efficiencies, number of expected background events,
and number of observed events for the six $\tau\tau$ channels. Only statistical uncertainties are included.
For the all-leptonic and semileptonic channels, numbers of expected
background events and observed events are evaluated for each mass point in \mzh~intervals
corresponding to $\pm$2.5 times the expected resolution.
For the all-hadronic channel we consider the number of expected background, signal, and observed events for $\mzh > 800$\GeV.
When the expected background is zero, the 68\% confidence level
upper limit is listed.}\label{tab:effbkg}
\footnotesize
\begin{tabular}{lc|cccccc}
\hline
& Mass (\TeVns{}) & $\te\te$ & $\te\tm$ & $\tm\tm$ & $\te\tauh$ & $\tm\tauh$ & $\tauh\tauh$ \\ \hline
$\mathcal{B}$($\tau\tau$) & & 3.2\% & 6.2\% & 3.0\% & 23.1\% & 22.6\% & 41.9\% \\ \hline
$\varepsilon_{\text{sig}} (\%)$ & 0.8 & 2.8 $\pm$ 0.7 & 3.4 $\pm$ 0.5 & 4.2 $\pm$ 0.7 & 3.3 $\pm$ 0.3 & 4.4 $\pm$ 0.3 & 2.2 $\pm$ 0.2 \\
                                  & 0.9 & 11 $\pm$ 1 & 16 $\pm$ 1 & 20 $\pm$ 2 & 14.3 $\pm$ 0.5 & 18.7 $\pm$ 0.6 & 11.5 $\pm$ 0.4 \\
                                  & 1.0 & 17 $\pm$ 2 & 24 $\pm$ 1 & 38 $\pm$ 2 & 21.2 $\pm$ 0.6 & 29.3 $\pm$ 0.7 & 18.0 $\pm$ 0.5 \\
                                  & 1.2 & 26 $\pm$ 2 & 30 $\pm$ 1 & 39 $\pm$ 2 & 28.3 $\pm$ 0.7 & 35.8 $\pm$ 0.7 & 23.0 $\pm$ 0.5 \\
                                  & 1.5 & 30 $\pm$ 2 & 42 $\pm$ 2 & 53 $\pm$ 2 & 29.2 $\pm$ 0.8 & 38.1 $\pm$ 0.9 & 29.1 $\pm$ 0.7 \\
                                  & 2.0 & 28 $\pm$ 2 & 39 $\pm$ 2 & 56 $\pm$ 3 & 31.1 $\pm$ 0.8 & 39.2 $\pm$ 0.9 & 31.9 $\pm$ 0.7 \\
                                  & 2.5 & 27 $\pm$ 2 & 37 $\pm$ 2 & 42 $\pm$ 2 & 26.8 $\pm$ 0.8 & 37.0 $\pm$ 0.8 & 30.1 $\pm$ 0.7 \\ \hline
$N_{\text{bkg}}$ & 0.8 & 0.3 $\pm$ 0.5 & 1.1  $\pm$ 0.8 & 1.6 $\pm$ 1.2  & 6.1 $\pm$ 2.0  & 6.7 $\pm$ 2.1 & \multirow{7}[0]{*}{$6.1^{+3.2}_{-2.5}$}  \\
                   & 0.9 & 0.5 $\pm$ 0.4 & 1.7  $\pm$ 1.2  & 3.8 $\pm$ 2.1 & 9.8 $\pm$ 3.2  & 9.2 $\pm$ 2.9 &  \\
                   & 1.0 & 1.4 $\pm$ 1.4 & 1.7  $\pm$ 1.0  & 2.0 $\pm$ 0.9 & 9.5 $\pm$ 3.5  & 7.6 $\pm$ 2.2 &  \\
                   & 1.2 & 1.2 $\pm$ 1.2 & 1.2  $\pm$ 0.8  & 1.4 $\pm$ 0.6 & 5.0 $\pm$ 2.0  & 6.6 $\pm$ 2.3 &  \\
                   & 1.5 & 0.4 $\pm$ 0.4 & 0.07 $\pm$ 0.04 & 0.9 $\pm$ 0.4 & 4.3 $\pm$ 1.8  & 2.6 $\pm$ 0.9 &  \\
                   & 2.0 & $<$0.5 & $<$0.4 & 0.7 $\pm$ 0.4 & 0.1  $\pm$ 0.1  & $<$0.4 &  \\
                   & 2.5 & $<$2.1 & $<$0.3 & 0.3 $\pm$ 0.1 & 0.18 $\pm$ 0.05 & $<$0.5 &  \\ \hline
$N_{\text{obs}}$ & 0.8 & 1 & 1 & 2 & 3 & 10 & \multirow{7}[0]{*}{8}  \\
                        & 0.9 & 2 & 2 & 3 & 4 & 13 &  \\
                        & 1.0 & 2 & 2 & 5 & 2 & 13 &  \\
                        & 1.2 & 0 & 1 & 3 & 5 & 12 &  \\
                        & 1.5 & 0 & 0 & 1 & 2 &  5 &  \\
                        & 2.0 & 0 & 1 & 0 & 0 &  0 &  \\
                        & 2.5 & 0 & 0 & 0 & 0 &  0 &  \\\hline
\end{tabular}
\end{table*}

\begin{figure}[!htb]
\centering
\includegraphics[width=\cmsFigWidth]{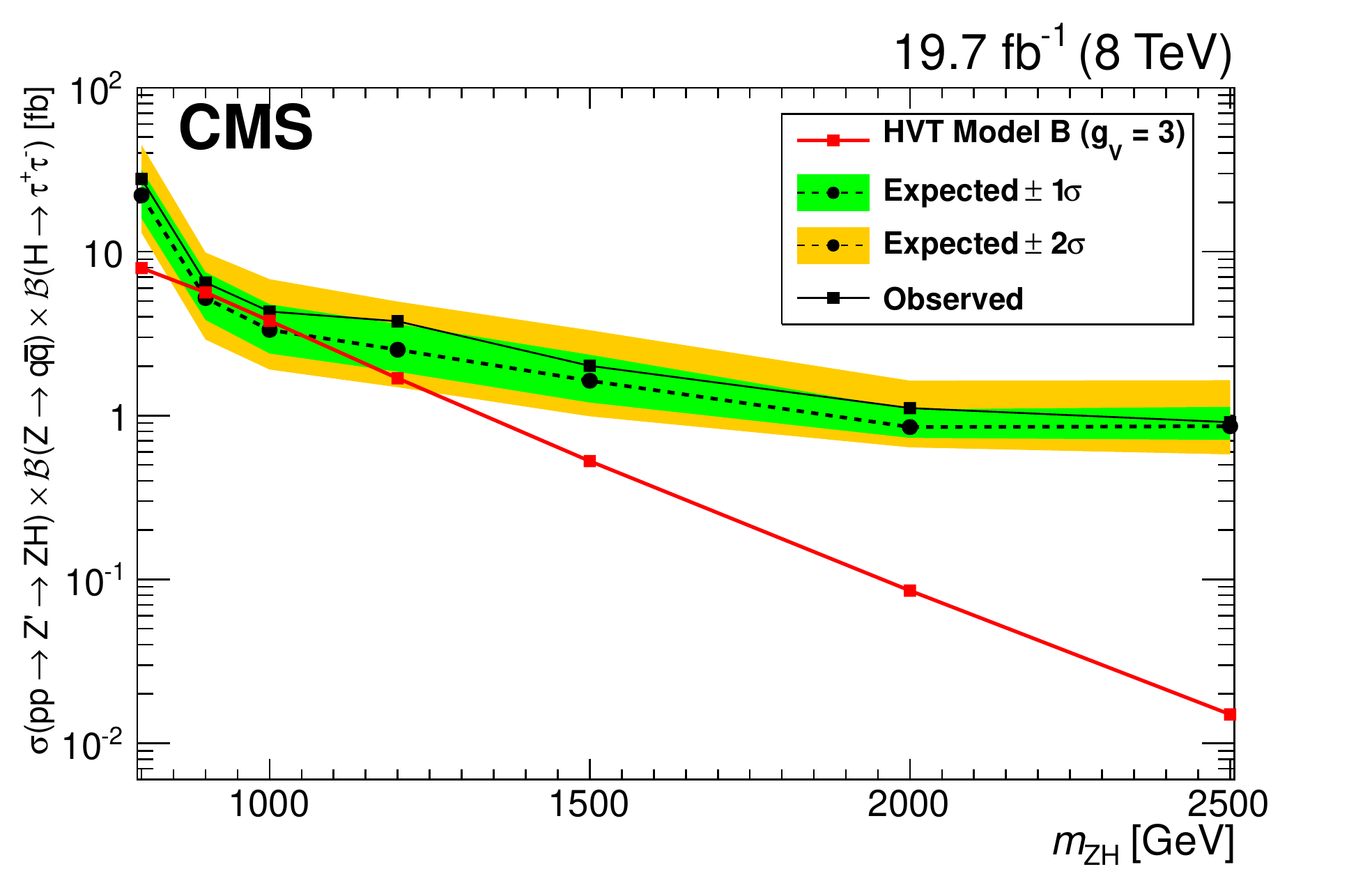}
\caption{Expected and observed upper limits on the quantity $\sigma(\PZpr) \: \mathcal{B}(\PZpr \to \Z\PH)$ for the six analysis channels combined.
Green and yellow bands correspond to $\pm$1 or $\pm2\sigma$ variations on the expected upper limit, respectively.}\label{fig:finallimit}

\end{figure}

Having observed no significant deviations in the observed number of events from the expected background,
we set upper limits on the production cross section of a new resonance in the ZH final state.
We use the CL$_\mathrm{s}$ criterion~\cite{CLs1,Junk:1999kv} to extract upper bounds on the cross section, combining
all six event categories. The test statistic is a profile likelihood ratio~\cite{ATL-PHYS-PUB-2011-011}
and the systematic uncertainties are
treated as nuisance parameters with the frequentist approach.
The nuisance parameters are described with log-normal prior probability distribution functions, except for
those related to the extrapolation from sideband events, which are expected to follow a $\Gamma$ distribution~\cite{ATL-PHYS-PUB-2011-011}. In the
all-leptonic and semileptonic channels, the numbers of signal and background events are
calculated for a region corresponding to $\pm$2.5 times the expected resolution around each mass point in \mzh,
while in the all-hadronic channel we consider the number of expected background, signal and observed events in $\mzh > 800$\GeV for each mass point.
The expected and observed upper limits are shown in Fig.~\ref{fig:finallimit}.
Production cross sections times branching fraction in a range between
0.9 and 27.8\unit{fb}, depending on the resonance mass (0.8--2.5\TeV), are
excluded at a 95\% confidence level.

\begin{figure}[!htb]
\centering
\includegraphics[width=\cmsFigWidth]{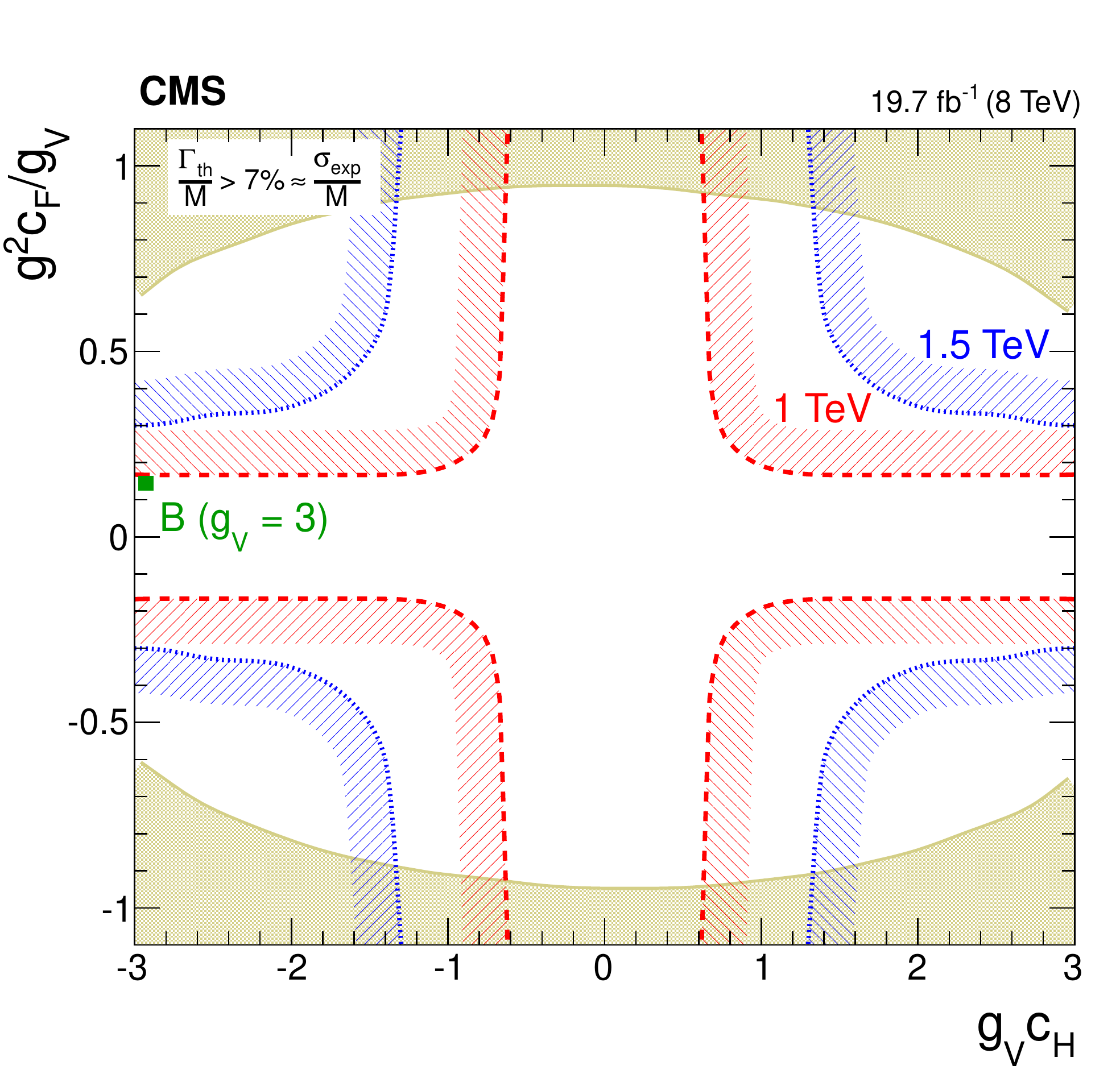}
\caption{Exclusion regions in the plane of the HVT-model coupling constants ($\givu\ciacca$,
$g^2c_\mathrm{F}/\givu$) for two resonance masses, 1.0 and 1.5\TeV.  The
point B of the benchmark model used in the analysis, corresponding to
$\givu=3$ and $c_\mathrm{F}=-\ciacca=1$, is also shown. The
boundaries of the regions of the plane excluded by this search are
indicated by the dashed and dotted lines, and associated hatching. The
areas indicated by the solid line and solid shading correspond to regions
where the theoretical width is larger than the experimental resolution of
the present search and thus the narrow-resonance assumption is not
satisfied.}\label{fig:coupling}

\end{figure}

In Fig.~\ref{fig:finallimit}, the results from this analysis are also compared to the cross section of the theoretical model, used as benchmark in this paper and studied in Ref.~\cite{Pappadopulo:2014qza}. In this model, the parameters are chosen to be $\givu=3$ and $c_\mathrm{F}=-\ciacca=1$,
corresponding to a strongly coupled sector. In Fig.~\ref{fig:coupling}, a scan of the coupling parameters and the corresponding regions of exclusion in the HVT model are shown. The parameters are defined as $\givu\ciacca$ and $g^2c_\mathrm{F}/\givu$, related to the coupling strength of the new resonance to the
Higgs boson and to fermions. Regions of the plane excluded by this search are indicated
by hatched
areas. Ranges of the scan are limited by the assumption
that the new resonance is narrow.

\section{Summary}

A search for a highly massive ($\geq$0.8\TeV) and
narrow resonance decaying to
Z and H bosons that decay in turn to merged dijet and $\tau^+\tau^-$ final states
has been conducted with data samples collected in 8\TeV proton-proton collisions by the CMS experiment in 2012.
For a high-mass resonance decaying to much lighter Z and H bosons, the final state
particles must be detected and reconstructed in small angular regions.  This is the first
search performed by adopting novel and advanced reconstruction techniques to
accomplish that end.
From a combination of all possible decay modes of the $\tau$ leptons,
production cross sections in a range between
0.9 and 27.8\unit{fb}, depending on the resonance mass (0.8--2.5\TeV), are
excluded at a 95\% confidence level.

\section*{Acknowledgements}
We congratulate our colleagues in the CERN accelerator departments for the excellent performance of the LHC and thank the technical and administrative staffs at CERN and at other CMS institutes for their contributions to the success of the CMS effort. In addition, we gratefully acknowledge the computing centers and personnel of the Worldwide LHC Computing Grid for delivering so effectively the computing infrastructure essential to our analyses. Finally, we acknowledge the enduring support for the construction and operation of the LHC and the CMS detector provided by the following funding agencies: BMWFW and FWF (Austria); FNRS and FWO (Belgium); CNPq, CAPES, FAPERJ, and FAPESP (Brazil); MES (Bulgaria); CERN; CAS, MoST, and NSFC (China); COLCIENCIAS (Colombia); MSES and CSF (Croatia); RPF (Cyprus); MoER, ERC IUT and ERDF (Estonia); Academy of Finland, MEC, and HIP (Finland); CEA and CNRS/IN2P3 (France); BMBF, DFG, and HGF (Germany); GSRT (Greece); OTKA and NIH (Hungary); DAE and DST (India); IPM (Iran); SFI (Ireland); INFN (Italy); MSIP and NRF (Republic of Korea); LAS (Lithuania); MOE and UM (Malaysia); CINVESTAV, CONACYT, SEP, and UASLP-FAI (Mexico); MBIE (New Zealand); PAEC (Pakistan); MSHE and NSC (Poland); FCT (Portugal); JINR (Dubna); MON, RosAtom, RAS and RFBR (Russia); MESTD (Serbia); SEIDI and CPAN (Spain); Swiss Funding Agencies (Switzerland); MST (Taipei); ThEPCenter, IPST, STAR and NSTDA (Thailand); TUBITAK and TAEK (Turkey); NASU and SFFR (Ukraine); STFC (United Kingdom); DOE and NSF (USA).

Individuals have received support from the Marie-Curie program and the European Research Council and EPLANET (European Union); the Leventis Foundation; the A. P. Sloan Foundation; the Alexander von Humboldt Foundation; the Belgian Federal Science Policy Office; the Fonds pour la Formation \`a la Recherche dans l'Industrie et dans l'Agriculture (FRIA-Belgium); the Agentschap voor Innovatie door Wetenschap en Technologie (IWT-Belgium); the Ministry of Education, Youth and Sports (MEYS) of the Czech Republic; the Council of Science and Industrial Research, India; the HOMING PLUS program of Foundation for Polish Science, cofinanced from European Union, Regional Development Fund; the Compagnia di San Paolo (Torino); the Consorzio per la Fisica (Trieste); MIUR project 20108T4XTM (Italy); the Thalis and Aristeia programs cofinanced by EU-ESF and the Greek NSRF; and the National Priorities Research Program by Qatar National Research Fund.
\bibliography{auto_generated}

\providecommand{\href}[2]{#2}\begingroup\raggedright\begin{thebibliography}{10}%
\makeatletter
\providecommand{\hrefCMSnoop }[0]{\@secondoftwo}%
\makeatother
\providecommand{\doi}{\texttt{doi:}\begingroup \urlstyle{tt}\Url}

\bibitem{Aad:2012tfa}
\hrefCMSnoop {}{{ATLAS} Collaboration, ``{Observation of a new particle in the
  search for the Standard Model Higgs boson with the ATLAS detector at the
  LHC}'',} \textit{ Phys. Lett. B} \textbf{ 716} (2012) 1,
  \href{http://dx.doi.org/10.1016/j.physletb.2012.08.020}{\doi{10.1016/j.physletb.2012.08.020}},
\href{http://www.arXiv.org/abs/1207.7214}{\texttt{arXiv:1207.7214}}.

\bibitem{HIGGS1}
\hrefCMSnoop {}{{CMS} Collaboration, ``{Observation of a new boson at a mass of
  125 GeV with the CMS experiment at the LHC}'',} \textit{ Phys. Lett. B}
  \textbf{ 716} (2012) 30,
  \href{http://dx.doi.org/10.1016/j.physletb.2012.08.021}{\doi{10.1016/j.physletb.2012.08.021}},
\href{http://www.arXiv.org/abs/1207.7235}{\texttt{arXiv:1207.7235}}.

\bibitem{ArkaniHamed:1998rs}
\hrefCMSnoop {}{N.~Arkani-Hamed, S.~Dimopoulos, and G.~R. Dvali, ``{The
  hierarchy problem and new dimensions at a millimeter}'',} \textit{ Phys.
  Lett. B} \textbf{ 429} (1998) 263,
  \href{http://dx.doi.org/10.1016/S0370-2693(98)00466-3}{\doi{10.1016/S0370-2693(98)00466-3}},
\href{http://www.arXiv.org/abs/hep-ph/9803315}{\texttt{arXiv:hep-ph/9803315}}.

\bibitem{PhysRevLett.49.970}
\hrefCMSnoop {}{A.~H. Chamseddine, R.~Arnowitt, and P.~Nath, ``Locally
  Supersymmetric Grand Unification'',} \textit{ Phys. Rev. Lett.} \textbf{ 49}
  (1982) 970,
  \href{http://dx.doi.org/10.1103/PhysRevLett.49.970}{\doi{10.1103/PhysRevLett.49.970}}.

\bibitem{Barbieri:1982eh}
\hrefCMSnoop {}{R.~Barbieri, S.~Ferrara, and C.~A. Savoy, ``{Gauge models with
  spontaneously broken local supersymmetry}'',} \textit{ Phys. Lett. B}
  \textbf{ 119} (1982) 343,
\href{http://dx.doi.org/10.1016/0370-2693(82)90685-2}{\doi{10.1016/0370-2693(82)90685-2}}.

\bibitem{Agashe:2004rs}
\hrefCMSnoop {}{K.~Agashe, R.~Contino, and A.~Pomarol, ``{The Minimal composite
  Higgs model}'',} \textit{ Nucl. Phys. B} \textbf{ 719} (2005) 165,
  \href{http://dx.doi.org/10.1016/j.nuclphysb.2005.04.035}{\doi{10.1016/j.nuclphysb.2005.04.035}},
\href{http://www.arXiv.org/abs/hep-ph/0412089}{\texttt{arXiv:hep-ph/0412089}}.

\bibitem{PhysRevD.75.055014}
\hrefCMSnoop {}{R.~Contino, L.~Da~Rold, and A.~Pomarol, ``Light custodians in
  natural composite {H}iggs models'',} \textit{ Phys. Rev. D} \textbf{ 75}
  (2007) 055014,
  \href{http://dx.doi.org/10.1103/PhysRevD.75.055014}{\doi{10.1103/PhysRevD.75.055014}}.

\bibitem{PhysRevLett.83.3370}
\hrefCMSnoop {}{L.~Randall and R.~Sundrum, ``Large Mass Hierarchy from a Small
  Extra Dimension'',} \textit{ Phys. Rev. Lett.} \textbf{ 83} (1999) 3370,
  \href{http://dx.doi.org/10.1103/PhysRevLett.83.3370}{\doi{10.1103/PhysRevLett.83.3370}}.

\bibitem{Zprime1}
\hrefCMSnoop {}{A.~Leike, ``{The phenomenology of extra neutral gauge
  bosons}'',} \textit{ Phys. Rept.} \textbf{ 317} (1999) 143,
  \href{http://dx.doi.org/10.1016/S0370-1573(98)00133-1}{\doi{10.1016/S0370-1573(98)00133-1}},
\href{http://www.arXiv.org/abs/hep-ph/9805494}{\texttt{arXiv:hep-ph/9805494}}.

\bibitem{Zprime2}
\hrefCMSnoop {}{T.~G. Rizzo, ``{$Z^\prime$ phenomenology and the LHC}'',}
  (2006).
\href{http://www.arXiv.org/abs/hep-ph/0610104}{\texttt{arXiv:hep-ph/0610104}}.

\bibitem{Composite1}
\hrefCMSnoop {}{R.~Contino, D.~Marzocca, D.~Pappadopulo, and R.~Rattazzi, ``{On
  the effect of resonances in composite Higgs phenomenology}'',} \textit{ JHEP}
  \textbf{ 10} (2011) 081,
  \href{http://dx.doi.org/10.1007/JHEP10(2011)081}{\doi{10.1007/JHEP10(2011)081}},
\href{http://www.arXiv.org/abs/1109.1570}{\texttt{arXiv:1109.1570}}.

\bibitem{Composite2}
\hrefCMSnoop {}{D.~Marzocca, M.~Serone, and J.~Shu, ``General composite {H}iggs
  models'',} \textit{ JHEP} \textbf{ 08} (2012) 013,
  \href{http://dx.doi.org/10.1007/JHEP08(2012)013}{\doi{10.1007/JHEP08(2012)013}},
\href{http://www.arXiv.org/abs/1205.0770}{\texttt{arXiv:1205.0770}}.

\bibitem{Aad2015163}
\hrefCMSnoop {}{{ATLAS} Collaboration, ``Search for a {CP}-odd {H}iggs boson
  decaying to {$Zh$} in $pp$ collisions at $\sqrt{s}=8$ {TeV} with the {ATLAS}
  detector'',} \textit{ Phys. Lett B} \textbf{ 744} (2015) 163,
  \href{http://dx.doi.org/10.1016/j.physletb.2015.03.054}{\doi{10.1016/j.physletb.2015.03.054}}.

\bibitem{Aad:2014pha}
\hrefCMSnoop {}{{ATLAS} Collaboration, ``{Search for $WZ$ resonances in the
  fully leptonic channel using $\rm{pp}$ collisions at $\sqrt{s}$ = 8 TeV with
  the ATLAS detector}'',} \textit{ Phys. Lett. B} \textbf{ 737} (2014) 223,
  \href{http://dx.doi.org/10.1016/j.physletb.2014.08.039}{\doi{10.1016/j.physletb.2014.08.039}},
\href{http://www.arXiv.org/abs/1406.4456}{\texttt{arXiv:1406.4456}}.

\bibitem{Aad:2014xka}
\hrefCMSnoop {}{{ATLAS} Collaboration, ``{Search for resonant diboson
  production in the $\mathrm {\ell \ell }q\bar{q}$ final state in $pp$
  collisions at $\sqrt{s} = 8$ TeV with the ATLAS detector}'',} \textit{ Eur.
  Phys. J. C} \textbf{ 75} (2015) 69,
  \href{http://dx.doi.org/10.1140/epjc/s10052-015-3261-8}{\doi{10.1140/epjc/s10052-015-3261-8}},
\href{http://www.arXiv.org/abs/1409.6190}{\texttt{arXiv:1409.6190}}.

\bibitem{Aad:2014yja}
\hrefCMSnoop {}{{ATLAS} Collaboration, ``Search For {H}iggs Boson Pair
  Production in the $\gamma\gamma b\bar{b}$ Final State using $pp$ Collision
  Data at $\sqrt{s}=8$ {TeV} from the {ATLAS} Detector'',} \textit{ Phys. Rev.
  Lett.} \textbf{ 114} (2015) 081802,
  \href{http://dx.doi.org/10.1103/PhysRevLett.114.081802}{\doi{10.1103/PhysRevLett.114.081802}},
\href{http://www.arXiv.org/abs/1406.5053}{\texttt{arXiv:1406.5053}}.

\bibitem{CMS:ZZWW}
\hrefCMSnoop {}{{CMS} Collaboration, ``{Search for massive resonances decaying
  into pairs of boosted bosons in semi-leptonic final states at $\sqrt{s} =$ 8
  TeV}'',} \textit{ JHEP} \textbf{ 08} (2014) 174,
  \href{http://dx.doi.org/10.1007/JHEP08(2014)174}{\doi{10.1007/JHEP08(2014)174}},
\href{http://www.arXiv.org/abs/1405.3447}{\texttt{arXiv:1405.3447}}.

\bibitem{Khachatryan:2014hpa}
\hrefCMSnoop {}{{CMS} Collaboration, ``{Search for massive resonances in dijet
  systems containing jets tagged as W or Z boson decays in pp collisions at $
  \sqrt{s} $ = 8 TeV}'',} \textit{ JHEP} \textbf{ 08} (2014) 173,
  \href{http://dx.doi.org/10.1007/JHEP08(2014)173}{\doi{10.1007/JHEP08(2014)173}},
\href{http://www.arXiv.org/abs/1405.1994}{\texttt{arXiv:1405.1994}}.

\bibitem{Khachatryan:2014xja}
\hrefCMSnoop {}{{CMS} Collaboration, ``{Search for new resonances decaying via
  WZ to leptons in proton-proton collisions at $\sqrt{s}$ = 8 TeV}'',} \textit{
  Phys. Lett. B} \textbf{ 740} (2014) 83,
  \href{http://dx.doi.org/10.1016/j.physletb.2014.11.026}{\doi{10.1016/j.physletb.2014.11.026}},
\href{http://www.arXiv.org/abs/1407.3476}{\texttt{arXiv:1407.3476}}.

\bibitem{RECENTAtlas}
\hrefCMSnoop {}{{ATLAS} Collaboration, ``{Search for a new resonance decaying
  to a W or Z boson and a Higgs boson in the $\ell\ell$/$\ell\nu$/$\nu\nu$ + $b
  {\bar b}$ final states with the ATLAS Detector}'',} \textit{ Eur. Phys. J. C}
  \textbf{ 75} (2015) 263,
  \href{http://dx.doi.org/10.1140/epjc/s10052-015-3474-x}{\doi{10.1140/epjc/s10052-015-3474-x}},
\href{http://www.arXiv.org/abs/1503.08089}{\texttt{arXiv:1503.08089}}.

\bibitem{Khachatryan:2015lba}
\hrefCMSnoop {}{{CMS Collaboration}, ``{Search for a pseudoscalar boson
  decaying into a Z boson and the 125 GeV Higgs boson in llbb final states}'',}
  (2015).
  \href{http://www.arXiv.org/abs/1504.04710}{\texttt{arXiv:1504.04710}}.
Submitted to PLB.

\bibitem{Khachatryan:2015yea}
\hrefCMSnoop {}{{CMS} Collaboration, ``{Search for resonant pair production of
  Higgs bosons decaying to two bottom quark-antiquark pairs in proton-proton
  collisions at 8 TeV}'',} (2015).
  \href{http://www.arXiv.org/abs/1503.04114}{\texttt{arXiv:1503.04114}}.
Submitted to PLB.

\bibitem{Aad:2015owa}
\hrefCMSnoop {}{{ATLAS Collaboration}, ``{Search for high-mass diboson
  resonances with boson-tagged jets in proton-proton collisions at $\sqrt{s}$ =
  8 TeV with the ATLAS detector}'',} (2015).
  \href{http://www.arXiv.org/abs/1506.00962}{\texttt{arXiv:1506.00962}}.
Submitted to JHEP.

\bibitem{Aad:2015uka}
\hrefCMSnoop {}{{ATLAS Collaboration}, ``{Search for Higgs boson pair
  production in the $b\bar{b} b\bar{b}$ final state from $pp$ collisions at
  $\sqrt{s} = 8$ TeV with the ATLAS detector}'',} (2015).
  \href{http://www.arXiv.org/abs/1506.00285}{\texttt{arXiv:1506.00285}}.
Submitted to EPJC.

\bibitem{Pappadopulo:2014qza}
\hrefCMSnoop {}{D.~Pappadopulo, A.~Thamm, R.~Torre, and A.~Wulzer, ``Heavy
  Vector Triplets: Bridging Theory and Data'',} (2014).
\href{http://www.arXiv.org/abs/1402.4431}{\texttt{arXiv:1402.4431}}.

\bibitem{Schmaltz:2010xr}
\hrefCMSnoop {}{M.~Schmaltz and C.~Spethmann, ``{Two simple W' models for the
  early LHC}'',} \textit{ JHEP} \textbf{ 07} (2011) 046,
  \href{http://dx.doi.org/10.1007/JHEP07(2011)046}{\doi{10.1007/JHEP07(2011)046}},
\href{http://www.arXiv.org/abs/1011.5918}{\texttt{arXiv:1011.5918}}.

\bibitem{Bellazzini:2012tv}
B.~Bellazzini\hrefCMSnoop {}{ {et~al.}, ``Composite {H}iggs sketch'',} \textit{
  JHEP} \textbf{ 11} (2012) 003,
  \href{http://dx.doi.org/10.1007/JHEP11(2012)003}{\doi{10.1007/JHEP11(2012)003}},
\href{http://www.arXiv.org/abs/1205.4032}{\texttt{arXiv:1205.4032}}.

\bibitem{Chatrchyan:2008zzk}
\hrefCMSnoop {}{{CMS} Collaboration, ``The {CMS} experiment at the {CERN}
  {LHC}'',} \textit{ JINST} \textbf{ 3} (2008) S08004,
\href{http://dx.doi.org/10.1088/1748-0221/3/08/S08004}{\doi{10.1088/1748-0221/3/08/S08004}}.

\bibitem{Cacciari:2008gp}
\hrefCMSnoop {}{M.~Cacciari, G.~P. Salam, and G.~Soyez, ``The anti-$k_t$ jet
  clustering algorithm'',} \textit{ JHEP} \textbf{ 04} (2008) 063,
  \href{http://dx.doi.org/10.1088/1126-6708/2008/04/063}{\doi{10.1088/1126-6708/2008/04/063}},
  \href{http://www.arXiv.org/abs/0802.1189}{\texttt{arXiv:0802.1189}}.

\bibitem{Alwall:2011uj}
J.~Alwall\hrefCMSnoop {}{ {et~al.}, ``{MadGraph} 5: going beyond'',} \textit{
  JHEP} \textbf{ 06} (2011) 128,
  \href{http://dx.doi.org/10.1007/JHEP06(2011)128}{\doi{10.1007/JHEP06(2011)128}},
\href{http://www.arXiv.org/abs/1106.0522}{\texttt{arXiv:1106.0522}}.

\bibitem{ref_SignalModel}
\href {{http://feynrules.irmp.ucl.ac.be/wiki/HiddenAbelianHiggsModel}}{C.~Duhr,
  ``{H}idden {A}belian {H}iggs {M}odel'',} 2011.

\bibitem{Nason:2004rx}
\hrefCMSnoop {}{P.~Nason, ``{A New method for combining NLO QCD with shower
  Monte Carlo algorithms}'',} \textit{ JHEP} \textbf{ 11} (2004) 040,
  \href{http://dx.doi.org/10.1088/1126-6708/2004/11/040}{\doi{10.1088/1126-6708/2004/11/040}},
\href{http://www.arXiv.org/abs/hep-ph/0409146}{\texttt{arXiv:hep-ph/0409146}}.

\bibitem{Frixione:2007vw}
\hrefCMSnoop {}{S.~Frixione, P.~Nason, and C.~Oleari, ``{Matching NLO QCD
  computations with parton shower simulations: the POWHEG method}'',} \textit{
  JHEP} \textbf{ 11} (2007) 070,
  \href{http://dx.doi.org/10.1088/1126-6708/2007/11/070}{\doi{10.1088/1126-6708/2007/11/070}},
\href{http://www.arXiv.org/abs/0709.2092}{\texttt{arXiv:0709.2092}}.

\bibitem{Alioli:2010xd}
\hrefCMSnoop {}{S.~Alioli, P.~Nason, C.~Oleari, and E.~Re, ``{A general
  framework for implementing NLO calculations in shower Monte Carlo programs:
  the POWHEG BOX}'',} \textit{ JHEP} \textbf{ 06} (2010) 043,
  \href{http://dx.doi.org/10.1007/JHEP06(2010)043}{\doi{10.1007/JHEP06(2010)043}},
\href{http://www.arXiv.org/abs/1002.2581}{\texttt{arXiv:1002.2581}}.

\bibitem{Alioli:2011as}
\hrefCMSnoop {}{S.~Alioli, S.-O. Moch, and P.~Uwer, ``{Hadronic top-quark
  pair-production with one jet and parton showering}'',} \textit{ JHEP}
  \textbf{ 01} (2012) 137,
  \href{http://dx.doi.org/10.1007/JHEP01(2012)137}{\doi{10.1007/JHEP01(2012)137}},
\href{http://www.arXiv.org/abs/1110.5251}{\texttt{arXiv:1110.5251}}.

\bibitem{Sjostrand:2006za}
\hrefCMSnoop {}{T.~Sj{\"o}strand, S.~Mrenna, and P.~Skands, ``{PYTHIA} 6.4
  physics and manual'',} \textit{ JHEP} \textbf{ 05} (2006) 026,
  \href{http://dx.doi.org/10.1088/1126-6708/2006/05/026}{\doi{10.1088/1126-6708/2006/05/026}},
\href{http://www.arXiv.org/abs/hep-ph/0603175}{\texttt{arXiv:hep-ph/0603175}}.

\bibitem{TAUOLA}
\hrefCMSnoop {}{S.~Jadach, J.~H. K{\"u}hn, and Z.~W{\c{a}}s, ``{TAUOLA} - a
  library of {M}onte {C}arlo programs to simulate decays of polarized tau
  leptons'',} \textit{ Comput. Phys. Commun.} \textbf{ 64} (1991) 275,
  \href{http://dx.doi.org/10.1016/0010-4655(91)90038-M}{\doi{10.1016/0010-4655(91)90038-M}}.

\bibitem{Agostinelli:2002hh}
\hrefCMSnoop {}{{GEANT4} Collaboration, ``{GEANT4}---a simulation toolkit'',}
  \textit{ Nucl. Instrum. Meth. A} \textbf{ 506} (2003) 250,
\href{http://dx.doi.org/10.1016/S0168-9002(03)01368-8}{\doi{10.1016/S0168-9002(03)01368-8}}.

\bibitem{CMS-PAS-PFT-09-001}
\href {http://cdsweb.cern.ch/record/1194487}{{CMS} Collaboration,
  ``Particle--Flow Event Reconstruction in {CMS} and Performance for Jets,
  Taus, and {\MET}'',} CMS Physics Analysis Summary CMS-PAS-PFT-09-001, CERN,
  2009.

\bibitem{CMS-PAS-PFT-10-001}
\href {http://cdsweb.cern.ch/record/1247373}{{CMS} Collaboration,
  ``Commissioning of the Particle-flow Event Reconstruction with the first
  {LHC} collisions recorded in the {CMS} detector'',} CMS Physics Analysis
  Summary CMS-PAS-PFT-10-001, 2010.

\bibitem{CMS-JME-10-011}
\hrefCMSnoop {}{{CMS} Collaboration, ``Determination of jet energy calibration
  and transverse momentum resolution in {CMS}'',} \textit{ JINST} \textbf{ 6}
  (2011) 11002,
  \href{http://dx.doi.org/10.1088/1748-0221/6/11/P11002}{\doi{10.1088/1748-0221/6/11/P11002}},
  \href{http://www.arXiv.org/abs/1107.4277}{\texttt{arXiv:1107.4277}}.

\bibitem{CMS-JME-12-002}
\href {http://cdsweb.cern.ch/record/1543527}{{CMS} Collaboration, ``Performance
  of Missing Transverse Momentum Reconstruction Algorithms in Proton-Proton
  Collisions at $\sqrt{s}=8$ {TeV} with the {CMS} Detector'',} CMS Physics
  Analysis Summary CMS-PAS-JME-12-002, CERN, 2012.

\bibitem{Wobisch:1998wt}
\hrefCMSnoop {}{M.~Wobisch and T.~Wengler, ``Hadronization corrections to jet
  cross-sections in deep inelastic scattering'',} (1998).
\href{http://www.arXiv.org/abs/hep-ph/9907280}{\texttt{arXiv:hep-ph/9907280}}.

\bibitem{Thaler:2010tr}
\hrefCMSnoop {}{J.~Thaler and K.~Van~Tilburg, ``{Identifying boosted objects
  with N-subjettiness}'',} \textit{ JHEP} \textbf{ 03} (2011) 015,
  \href{http://dx.doi.org/10.1007/JHEP03(2011)015}{\doi{10.1007/JHEP03(2011)015}},
\href{http://www.arXiv.org/abs/1011.2268}{\texttt{arXiv:1011.2268}}.

\bibitem{Catani:1993hr}
\hrefCMSnoop {}{S.~Catani, Y.~L. Dokshitzer, M.~H. Seymour, and B.~R. Webber,
  ``{Longitudinally invariant anti-$k_{t}$ clustering algorithms for
  hadron-hadron collisions}'',} \textit{ Nucl. Phys. B} \textbf{ 406} (1993)
  187,
  \href{http://dx.doi.org/10.1016/0550-3213(93)90166-M}{\doi{10.1016/0550-3213(93)90166-M}}.

\bibitem{Ellis:1993tq}
\hrefCMSnoop {}{S.~D. Ellis and D.~E. Soper, ``{Successive combination jet
  algorithm for hadron collisions}'',} \textit{ Phys. Rev. D} \textbf{ 48}
  (1993) 3160,
  \href{http://dx.doi.org/10.1103/PhysRevD.48.3160}{\doi{10.1103/PhysRevD.48.3160}},
  \href{http://www.arXiv.org/abs/hep-ph/9305266}{\texttt{arXiv:hep-ph/9305266}}.

\bibitem{Khachatryan:2014vla}
\hrefCMSnoop {}{{CMS} Collaboration, ``{Identification techniques for highly
  boosted W bosons that decay into hadrons}'',} \textit{ JHEP} \textbf{ 12}
  (2014) 017,
  \href{http://dx.doi.org/10.1007/JHEP12(2014)017}{\doi{10.1007/JHEP12(2014)017}},
\href{http://www.arXiv.org/abs/1410.4227}{\texttt{arXiv:1410.4227}}.

\bibitem{Baffioni:2006cd}
S.~Baffioni\hrefCMSnoop {}{ {et~al.}, ``Electron reconstruction in {CMS}'',}
  \textit{ Eur. Phys. J. C} \textbf{ 49} (2007) 1099,
\href{http://dx.doi.org/10.1140/epjc/s10052-006-0175-5}{\doi{10.1140/epjc/s10052-006-0175-5}}.

\bibitem{vertexes}
\hrefCMSnoop {}{{CMS} Collaboration, ``{Description and performance of track
  and primary-vertex reconstruction with the CMS tracker}'',} \textit{ JINST}
  \textbf{ 9} (2014) P10009,
  \href{http://dx.doi.org/10.1088/1748-0221/9/10/P10009}{\doi{10.1088/1748-0221/9/10/P10009}},
\href{http://www.arXiv.org/abs/1405.6569}{\texttt{arXiv:1405.6569}}.

\bibitem{Chatrchyan:2012xi}
\hrefCMSnoop {}{{CMS} Collaboration, ``{Performance of CMS muon reconstruction
  in pp collision events at $\sqrt{s}=7$ TeV}'',} \textit{ JINST} \textbf{ 7}
  (2012) P10002,
  \href{http://dx.doi.org/10.1088/1748-0221/7/10/P10002}{\doi{10.1088/1748-0221/7/10/P10002}},
\href{http://www.arXiv.org/abs/1206.4071}{\texttt{arXiv:1206.4071}}.

\bibitem{CMS-PAS-TAU-11-001}
\hrefCMSnoop {}{{CMS} Collaboration, ``{Performance of tau-lepton
  reconstruction and identification in CMS}'',} \textit{ JINST} \textbf{ 7}
  (2012) P01001,
  \href{http://dx.doi.org/10.1088/1748-0221/7/01/P01001}{\doi{10.1088/1748-0221/7/01/P01001}},
\href{http://www.arXiv.org/abs/1109.6034}{\texttt{arXiv:1109.6034}}.

\bibitem{Butterworth:2008iy}
\hrefCMSnoop {}{J.~M. Butterworth, A.~R. Davison, M.~Rubin, and G.~P. Salam,
  ``{Jet substructure as a new Higgs search channel at the LHC}'',} \textit{
  Phys. Rev. Lett.} \textbf{ 100} (2008) 242001,
  \href{http://dx.doi.org/10.1103/PhysRevLett.100.242001}{\doi{10.1103/PhysRevLett.100.242001}},
\href{http://www.arXiv.org/abs/0802.2470}{\texttt{arXiv:0802.2470}}.

\bibitem{Chatrchyan:2014nva}
\hrefCMSnoop {}{{CMS} Collaboration, ``{Evidence for the 125 GeV Higgs boson
  decaying to a pair of $\tau$ leptons}'',} \textit{ JHEP} \textbf{ 05} (2014)
  104,
  \href{http://dx.doi.org/10.1007/JHEP05(2014)104}{\doi{10.1007/JHEP05(2014)104}},
\href{http://www.arXiv.org/abs/1401.5041}{\texttt{arXiv:1401.5041}}.

\bibitem{Chatrchyan:2012jua}
\hrefCMSnoop {}{{CMS} Collaboration, ``{Identification of b-quark jets with the
  CMS experiment}'',} \textit{ JINST} \textbf{ 8} (2013) P04013,
  \href{http://dx.doi.org/10.1088/1748-0221/8/04/P04013}{\doi{10.1088/1748-0221/8/04/P04013}},
\href{http://www.arXiv.org/abs/1211.4462}{\texttt{arXiv:1211.4462}}.

\bibitem{punzi}
\hrefCMSnoop {}{G.~Punzi, ``Sensitivity of searches for new signals and its
  optimization'',} in \textit{ PhyStat2003: Statistical Problems in Particle
  Physics, Astrophysics, and Cosmology}, L.~Lyons, R.~P. Mount, and
  R.~Reitmeyer, eds.
\newblock 2003.
\newblock
  \href{http://www.arXiv.org/abs/physics/0308063v2}{\texttt{arXiv:physics/0308063v2}}.

\bibitem{LumiUncert}
\href {http://cdsweb.cern.ch/record/1598864}{{CMS} Collaboration, ``CMS
  Luminosity Based on Pixel Cluster Counting - Summer 2013 Update'',} CMS
  Physics Analysis Summary CMS-PAS-LUM-13-001, 2013.

\bibitem{CMS-PAS-EGM-10-004}
\href {http://cdsweb.cern.ch/record/1299116}{{CMS} Collaboration, ``Electron
  Reconstruction and Identification at $\sqrt{s} = 7$ {TeV}'',} CMS Physics
  Analysis Summary CMS-PAS-EGM-10-004, CERN, 2010.

\bibitem{CMS-JME-13-006}
\href {http://cdsweb.cern.ch/record/1577417}{{CMS} Collaboration, ``Identifying
  Hadronically Decaying Vector Bosons Merged into a Single Jet'',} CMS Physics
  Analysis Summary CMS-PAS-JME-13-006, CERN, 2013.

\bibitem{CLs1}
\hrefCMSnoop {}{A.~L. Read, ``Presentation of search results: The {$CL_{s}$}
  technique'',} \textit{ J. Phys. G} \textbf{ 28} (2002) 2693,
\href{http://dx.doi.org/10.1088/0954-3899/28/10/313}{\doi{10.1088/0954-3899/28/10/313}}.

\bibitem{Junk:1999kv}
\hrefCMSnoop {}{T.~Junk, ``Confidence level computation for combining searches
  with small statistics'',} \textit{ Nucl. Instrum. Meth. A} \textbf{ 434}
  (1999) 435,
  \href{http://dx.doi.org/10.1016/S0168-9002(99)00498-2}{\doi{10.1016/S0168-9002(99)00498-2}},
\href{http://www.arXiv.org/abs/hep-ex/9902006}{\texttt{arXiv:hep-ex/9902006}}.

\bibitem{ATL-PHYS-PUB-2011-011}
\href {https://cds.cern.ch/record/1379837}{{ATLAS and CMS Collaborations},
  ``{Procedure for the LHC Higgs boson search combination in summer 2011}'',}
  Technical Report ATL-PHYS-PUB-2011-011, CMS-NOTE-2011-005, CERN, 2011.

\end{thebibliography}\endgroup

\cleardoublepage \appendix\section{The CMS Collaboration \label{app:collab}}\begin{sloppypar}\hyphenpenalty=5000\widowpenalty=500\clubpenalty=5000\textbf{Yerevan Physics Institute,  Yerevan,  Armenia}\\*[0pt]
V.~Khachatryan, A.M.~Sirunyan, A.~Tumasyan
\vskip\cmsinstskip
\textbf{Institut f\"{u}r Hochenergiephysik der OeAW,  Wien,  Austria}\\*[0pt]
W.~Adam, T.~Bergauer, M.~Dragicevic, J.~Er\"{o}, M.~Friedl, R.~Fr\"{u}hwirth\cmsAuthorMark{1}, V.M.~Ghete, C.~Hartl, N.~H\"{o}rmann, J.~Hrubec, M.~Jeitler\cmsAuthorMark{1}, W.~Kiesenhofer, V.~Kn\"{u}nz, M.~Krammer\cmsAuthorMark{1}, I.~Kr\"{a}tschmer, D.~Liko, I.~Mikulec, D.~Rabady\cmsAuthorMark{2}, B.~Rahbaran, H.~Rohringer, R.~Sch\"{o}fbeck, J.~Strauss, W.~Treberer-Treberspurg, W.~Waltenberger, C.-E.~Wulz\cmsAuthorMark{1}
% ...
\vskip\cmsinstskip
% ...
1:~~Also at Vienna University of Technology, Vienna, Austria\\
2:~~Also at CERN, European Organization for Nuclear Research, Geneva, Switzerland\\
% ...

\end{sloppypar}
\end{document}